\documentclass[12pt]{article}
\usepackage[normalem]{ulem}
\usepackage{soul}
\usepackage{verbatim,color,amssymb}
\usepackage{algorithm}
\usepackage{algpseudocode}
\usepackage{amsmath}
\usepackage{bbm}
\usepackage{amsthm}					
\usepackage{natbib}
\usepackage{multirow}
 \usepackage{makecell}
\usepackage{setspace}
\usepackage[mathscr]{euscript}
\usepackage{fancyhdr}
\usepackage{enumitem}
\usepackage{graphicx}
\usepackage{subcaption}
\usepackage{lineno}
\usepackage{array,booktabs}
\usepackage{url}
\usepackage{caption}
\usepackage{xcolor}
\usepackage{multibib}
\newcites{latex}{References}

\usepackage{lscape}

\usepackage{tikz}
\usetikzlibrary{arrows}
\usepackage{multirow}

\newtheorem*{Proof*}{Proof}

\def\I{{\cal I}}

\def\Ind{\hbox{I}}

\def\log{\hbox{log}}

\def\Bernoulli{\hbox{Bernoulli}}

\def\Beta{\hbox{Beta}}

\def\Exp{\hbox{Exp}}

\def\IG{\hbox{Inv-Ga}}

\def\Normal{\hbox{Normal}}

\def\Unif{\hbox{Unif}}

\def\bse{\begin{eqnarray*}}
\def\ese{\end{eqnarray*}}
\def\be{\begin{eqnarray}}
\def\ee{\end{eqnarray}}
\def\bq{\begin{equation}}
\def\eq{\end{equation}}

\usepackage{verbatim,color,amssymb}

\def\bone{{\mathbf 1}}

\def\b1e{{\mathbf e}}

\def\bx{{\mathbf I}}

\def\bq{{\mathbf q}}

\def\bt{{\mathbf t}}

\def\bw{{\mathbf w}}

\def\bx{{\mathbf x}}
\def\bX{{\mathbf X}}

\def\bz{{\mathbf z}}
\def\bZ{{\mathbf Z}}

\def\bzero{{\mathbf 0}}

\newcommand{\bmu}{\mbox{\boldmath $\mu$}}

\newcommand{\btheta}{\mbox{\boldmath $\theta$}}
\newcommand{\bbeta}{\mbox{\boldmath $\beta$}}

\newcommand{\bzeta}{\mbox{\boldmath $\zeta$}}
\newcommand{\bsigma}{\mbox{\boldmath $\sigma$}}

\newcommand{\blambda}{\mbox{\boldmath $\lambda$}}
\newcommand{\bLambda}{\mbox{\boldmath $\Lambda$}}

\newcommand{\bGamma}{\mbox{\boldmath $\Gamma$}}

\def\bz{{\mathbf z}}
\def\bI{{\mathbf I}}

\renewcommand\footnoterule{\kern-3pt \hrule \textwidth 2in \kern 2.6pt}

\def\boxit#1{\vbox{\hrule\hbox{\vrule\kern6pt \vbox{\kern6pt \textcolor{blue}{#1}\kern6pt}\kern6pt\vrule}\hrule}}

\def\authorfootnote#1{{\let\thefootnote\relax\footnotetext{#1}}}

\allowdisplaybreaks

\title{Bayesian Neural-Net-Assisted Multi-Treatment Mixture Cure Survival Model with Application in Pediatric Oncology}
\author{Peter Chang$^1$, John Kairalla$^1$, Arkaprava Roy$^1$\\
{\it $^1$Department of Biostatistics, University of Florida, Gainesville, USA}\\
Correspondence: arkaprava.roy@ufl.edu}
\date{}
\begin{document}
\maketitle
\thispagestyle{empty}
\baselineskip=24pt

\begin{abstract}
{Estimating covariate-conditional treatment effects in multi-arm oncology studies is complicated when treatment arms have common distributional features and a non-negligible fraction of patients achieve long-term remission. We propose a joint mixture cure model with covariate-dependent mixtures of log-normal kernels with treatment-specific inclusions. Both linear and neural-network-assisted non-linear covariate links are proposed. Specifically, the susceptible survival distributions use a common finite dictionary of log-normal components, and then a binary inclusion matrix $\bGamma$ determines which components are active in each treatment arm. All parameters, including the hidden bases of the neural network, are learned jointly, while the output coefficients remain treatment- or component-specific. Posterior inference is performed using gradient-based MCMC, and treatment effects are summarized by covariate-conditional differences in restricted mean survival time (RMST). Variable importance is assessed using thresholded marginal best linear projections with data partitioning. Across two simulation settings, the proposed method demonstrates good finite-sample performance, with lower RMST-based estimation error than \texttt{flexsurvcure}. Compared with pairwise \texttt{grf} fits, the proposed method yields lower RMST-contrast MSE in most comparisons while ensuring mutually coherent multi-treatment contrasts. Finally, the application to the AALL0434 trial reveals covariate-dependent patterns in RMST posterior across methotrexate-based regimens and provides new insights into how these differences vary with patient covariates, highlighting the method's practical utility for studying heterogeneous treatment effects in pediatric oncology trials.}

\vspace{1em}\noindent {\it Keywords:} Bayesian inference;
Cure models;
Heterogeneous treatment effects;
Mixture models;
Multi-treatment studies;
Neural networks;
Restricted mean survival time.
\end{abstract}
\newpage
\section{Introduction} \label{introduction}

{Treatment effects often vary across patients with different clinical characteristics. The average treatment effect, the population-level mean difference in outcomes under two treatments, can obscure clinically important variation across such patient characteristics. This is particularly important in oncology, where disease-free and overall survival are common time-to-event endpoints, and survival probabilities and restricted mean survival time (RMST) summarize their distributions \citep{powers2018some,xu2024estimating,hu2021estimating}. Let $T^{(g)}$ denote the potential event time under treatment arm $g$. For a prespecified horizon $h$, we can define the covariate-conditional RMST treatment effect as $\zeta^{(h)}_{gg'}(\bx)=\mathbb{E}\left[\min{T^{(g)},h}-\min{T^{(g')},h}\mid\bX=\bx\right]$. This estimand is the average treatment effect among patients with covariates $\bx$. 

Estimating such covariate-conditional RMST effects in multi-arm studies requires balancing information sharing with arm-specific modeling: separate fits may lose precision, whereas complete pooling removes treatment differences. AALL0434 illustrates this setting through four regimens: escalating-dose methotrexate with pegaspargase (C-MTX) and high-dose methotrexate (HDMTX), each with or without nelarabine. All regimens share the augmented Berlin--Frankfurt--M\"unster backbone, and some pairs also share the methotrexate strategy or nelarabine exposure \citep{dunsmore2020children}. These commonalities motivate adaptive borrowing, although the learned mixture components represent shared survival features rather than particular drugs or mechanisms. Our goal is therefore to learn cross-arm sharing while preserving arm-specific survival distributions and accommodating a cure fraction.
}



Mixture cure models offer a principled approach to account for long-term survivors by decomposing the population into cured and susceptible subgroups \citep{boag1949maximum, berkson1952survival}. Estimation of mixture cure models has been explored in both frequentist and Bayesian settings. In the frequentist framework, the expectation-maximization (EM) algorithm is commonly used for mixture models to help estimate the latent cure status \citep{kuk1992mixture, zhang2007new, peng2011mixture}, while Bayesian approaches rely on Markov chain Monte Carlo (MCMC) methods with data augmentation \citep{kuo2000mixture, de2022bayesian, pan2024bayesian}. 
{Promotion-time cure models provide an alternative cure formulation; common regression specifications impose a proportional-hazards structure on the marginal survival distribution \citep{chen1999new,ibrahim2001bayesian,yin2005cure}.}
However, traditional data-augmentation-based MCMC can be difficult to implement when jointly modeling cure status and heterogeneous treatment effects.
We thus avoid data augmentation and instead consider modern gradient-based MCMC methods. In mixture cure settings, where treatment may affect both cure probability and survival among susceptible patients, RMST summarizes differences in the full survival distribution up to a prespecified horizon without requiring proportional hazards.


{Our goal is to estimate and predict covariate-conditional RMST contrasts across multiple treatment arms. We propose a Bayesian multi-treatment mixture cure model with adaptive sharing. A common set of log-normal components represents susceptible survival, with $\bGamma$ selecting active components in each arm. This enables partial pooling, while treatment-specific cure functions and mixture weights preserve arm-level differences. Covariate effects use linear or single-layer neural-network links with shared hidden features. Posterior computation combines gradient-based MCMC with Gibbs updates for $\bGamma$. Joint estimation yields compatible RMST contrasts, whose covariate associations are summarized by data-partitioned marginal best linear projections and a threshold-indexed importance measure \citep{semenova2021debiased,cui2023estimating,chang2025}.}


\section{AALL0434 dataset}  \label{aall}
{The motivating data, analyzed in this paper, is from AALL0434, a phase III randomized trial conducted by the Children's Oncology Group among children and young adults with newly diagnosed T-cell acute lymphoblastic leukemia (T-ALL) or T-cell lymphoblastic lymphoma (T-LL) \citep{dunsmore2020children}. The trial enrolled 1,895 patients from 2007 to 2014. Age, white blood cell (WBC) count, sex, and central nervous system (CNS) status were recorded at diagnosis. Risk group was assigned after induction using clinical features and early response, including end-induction minimal residual disease.}

{After common induction and risk classification, the trial compared C-MTX with HDMTX and, among eligible patients, nelarabine with no nelarabine in a $2\times2$ pseudo-factorial design. Low-risk patients were not eligible for nelarabine, while CNS3 disease and induction failure led to additional assignment restrictions. These rules produced unequal overall arm sizes. The trial was not a SMART design, and cranial radiation was protocol-directed rather than randomized.} Excluding these cases yields a final cohort of 1,022 patients.
We provide a demographics table of the study population in Table S5. 

As an exploratory analysis, we plot the Kaplan-Meier curves for each treatment arm in Figure S2. Additionally, we also plot the Kaplan-Meier curves stratifying by key categorical covariates in Figures S3-S7.
We see that all tails of the curves have a plateau, indicating that a cure model may be more appropriate to use \citep{othus2012cure}.

\section{{Covariate-dependent mixture cure survival model}} \label{sec:model}

In standard survival analysis, it is often assumed that all individuals are susceptible to the event of interest and will eventually experience it, given sufficient follow-up. However, in many clinical contexts, a subset of patients may be considered cured and no longer at risk of the event. Following \cite{peng2000nonparametric} and \cite{li2007mixture}, we characterize the overall survival probability as a mixture model. 

{Let $T$ denote the event time and $B=I(T<\infty)$ the susceptibility indicator, with $B=1$ for a susceptible individual and $B=0$ for a cured individual.}
{For a patient with covariates $\bx_i$ under treatment arm $g$, the survival function is} $
    S_g(t \mid \bx_i) = c_g(\bx_i) + \{1-c_g(\bx_i)\} S_{g,u}(t \mid \bz_i)$
{Here, $c_g(\bx_i)=P(B_i=0\mid\bX=\bx_i,G=g)$ is the cure probability under treatment arm $g$, and $S_{g,u}(t\mid\bz_i)=P(T>t\mid\bZ=\bz_i,B=1,G=g)$ is the treatment-specific survival function among susceptible individuals.}
Our covariate matrices $\bX$ and $\bZ$ reflect covariates that can affect the two groups, respectively. 
In this paper, we consider $\bz_{i}=\bx_i$ i.e. $\bZ$ is identical to $\bX$ \citep{sy2000estimation, lee2024partly, cai2012smcure} such that the same covariates affect both the cured and susceptible patients. We characterize $c_{g}(\bx_{i})$ as,
\begin{align}
    c_g(\bx_i) = \frac{\exp(f_g^{(\text{cure})}(\bx_{i}))}{1 + \exp(f_g^{(\text{cure})}(\bx_{i}))} = \frac{1}{1 + \exp(-f_g^{(\text{cure})}(\bx_{i}))},\label{eq:cureprob}
\end{align}
{We specify $f_g^{(\text{cure})}(\cdot)$ using either the linear or neural-network covariate link described below.}

Next, we discuss our model for the survival probability, $S_{g,u}(t \mid \bx_{i})$, of the susceptible subjects. Here, we consider a covariate-dependent {treatment-specific} mixture of log-normals and let
$S_{g,u}(t \mid \bx_{i}, \bmu, \bsigma) = \sum_{m=1}^M \pi_{i,m,g} Q_{m}(t),$ 
where $Q_{m}(t)=P_{(\mu_m,\sigma_m)}(T>t)$ under a univariate {log-normal} distribution with parameters $\mu_m$ and $\sigma_{m}$.
The weights $\pi_{i,m,g}$ are modeled using a softmax output
function  
{
\begin{align}
    \pi_{i,m,g} = \frac{\gamma_{m,g}\exp\{f_{m}^{(\text{surv})}(\bx_{i})\}}{\sum_{j=1}^M \gamma_{j,g}\exp\{f_{j}^{(\text{surv})}(\bx_{i})\}},\label{eq:weights}
\end{align}
}
where the parameter $\bGamma=(\!(\gamma_{m,g})\!)_{1\leq m\leq M, 1\leq g\leq G}$ is a $M\times G$ binary matrix. {If $\gamma_{m,g}=\gamma_{m,g'}=1$, component $m$ is available in both arms and its location and fixed scale are shared; its normalized weight can still differ when the arms have different active component sets. A block-diagonal $\bGamma$ gives disjoint component sets and T-learning-type fits. If every entry equals one, the susceptible survival distribution is fully pooled across arms, while the cure functions remain treatment-specific. We estimate $\bGamma$ as a parameter that allows data-driven learning of an intermediate degree of sharing.} In the following subsections, we propose two possible choices for 
$f_{g}^{(\text{cure})}$'s and $f_{m}^{(\text{surv})}$'s.
We call them the {covariate} link functions.
Motivated by our application, we next further discuss two forms for our functions: a linear form and a nonlinear neural network form.

\subsection{{Linear Covariate} Link}
\begin{figure}[!htpb]
    \centering
    \includegraphics[width = 1\linewidth]{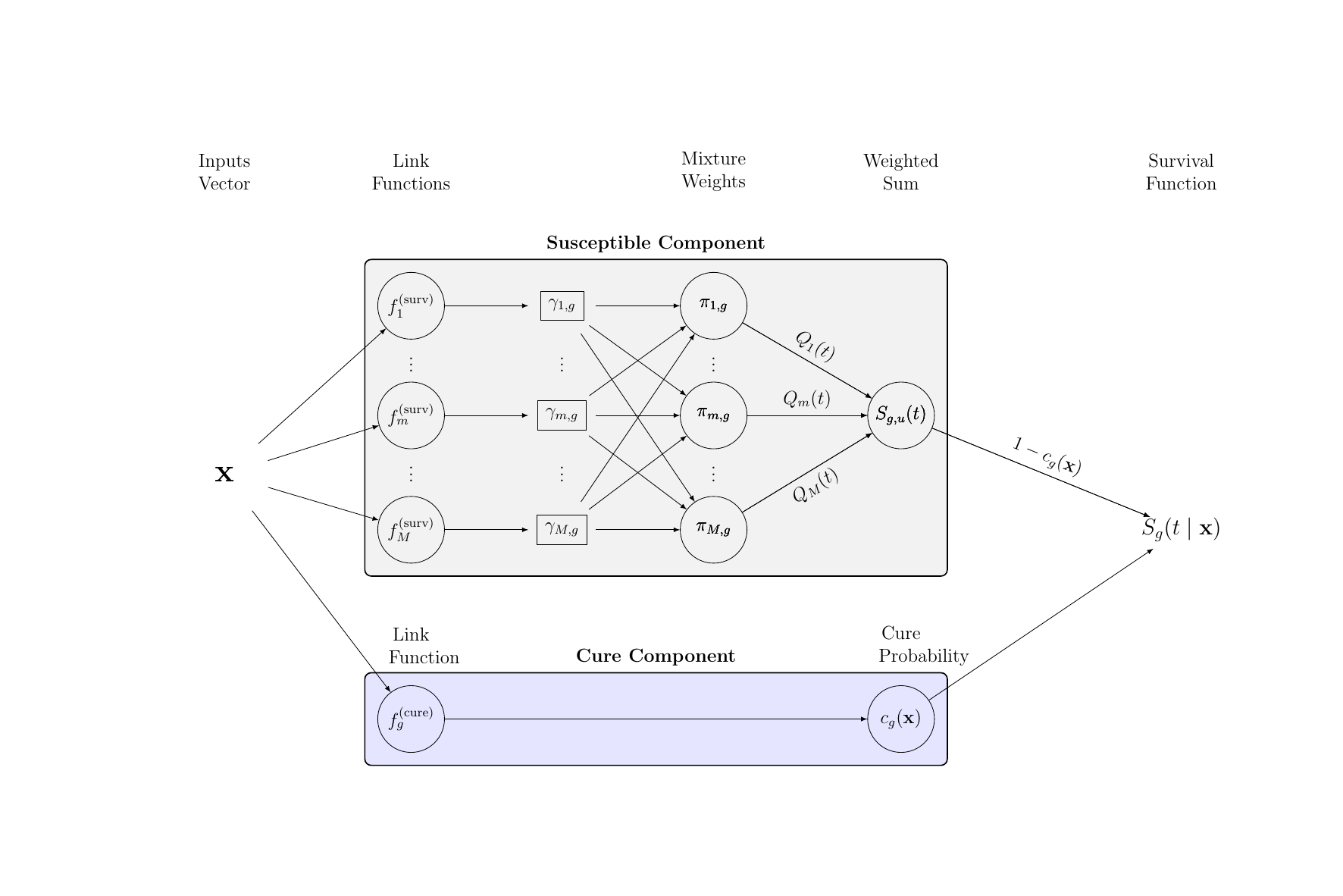}
    \caption{{Schematic of the proposed multi-treatment mixture cure model. The shared hidden units feed treatment-specific cure links and component-specific mixture-weight links. The indicators $\gamma_{m,g}$ determine which susceptible-survival components are active in each arm. Arrows show the direction from covariates through the link functions and mixture weights to $S_g(t\mid\bx_i)$.}}
    \label{fig:mcc_schematic}
\end{figure}

We first consider linear link functions, with
$f_g^{(\text{cure})}(\bx_i)=\bx_i^T\blambda_g
=\lambda_{g,0}+\sum_{p=1}^P x_{i,p}\lambda_{g,p}$ and
$f_m^{(\text{surv})}(\bx_i)=\bx_i^T\bbeta_m
=\beta_{m,0}+\sum_{p=1}^P x_{i,p}\beta_{m,p}$, where $x_{i,0}=1$, so
$\lambda_{g,0}$ and $\beta_{m,0}$ are intercepts. For $p\geq1$,
$\lambda_{g,p}$ represents the covariate effect on the cure linear predictor
for arm $g$, while $\beta_{m,p}$ has an analogous role for the weight of
component $m$. Since additive linear predictors cannot capture nonlinearities
or interactions, we also introduce a neural-network basis expansion for more
flexible covariate effects in the next subsection. Figure~\ref{fig:mcc_schematic} provides a schematic
of the model.

\subsection{Neural Network (NN) {Covariate} Link}
{We model both covariate links using a flexible neural network parametrization.} Specifically, we use the NN architecture and let $f_{g}^{(\text{cure})}(\bx_{i})=\lambda_{g,0}+\sum_{k=1}^K \lambda_{g,k}\psi(\bx_i^T\btheta_{k})$ and $f_{m}^{(\text{surv})}(\bx_{i})=\beta_{m,0}+\sum_{k=1}^K \beta_{m,k}\psi(\bx_i^T\btheta_{k})$, where
the activation function is set as the hyperbolic tangent function $\psi(a)=\frac{e^{2a}-1}{e^{2a}+1}$. {The hyperbolic tangent is used only in the hidden layer; the final cure probability is obtained by the logistic transformation in Equation~\eqref{eq:cureprob}. Shared hidden units capture common nonlinear structure, while link-specific output weights allow overall distinct covariate effects.}

Our choice of $K$ is further explained in Section \ref{specifyK}, where we discuss a cross-validation method balancing model complexity and performance. 
Early results from \cite{hornik1989multilayer, cybenko1989approximation} establish that a single hidden layer with a sufficient number of neurons possesses universal approximation properties, allowing our NN link functions to reflect {nonlinear} effects of the covariates \citep{anastassiou2011multivariate, xu2023bayesian}. 
In our mixture-cure setting, sharing both the neurons and activation functions across all the {covariate} links $f_{m}^{(\text{surv})}(\bx)$'s and $f_{g}^{(\text{cure})}(\bx)$ with function-specific weights ensures complex dependencies are learned jointly.




{\paragraph{Interpretation and assumptions}
The identification of $\zeta_{gg'}^{(h)}(\bx)$ as the covariate-conditional treatment effect of $g$ over $g'$ for an array of covariates $\bx$ requires consistency and no interference, treatment positivity, treatment exchangeability, and independent censoring with adequate support through $h$. In randomized trials, exchangeability follows from randomization within the population eligible for the comparison; in observational studies, it must hold conditional on the measured covariates. Under these conditions, the treatment-specific survival curves on $[0,h]$ and their RMST contrasts are identified. In AALL0434, the 10-year RMST horizon lies within approximately 12.5 years of follow-up, with patients remaining at risk in every arm at 10 years. The Kaplan--Meier curves also show clear tail flattening, consistent with clinically plausible long-term remission. We therefore retain $c_g(\bx)$ as a direct and parsimonious representation of this plateau, but infer on the marginal model.}

\section{Bayesian Inference}  \label{bayesian}

{We pursue a Bayesian approach to carry out the inference of the proposed cure mixture model and put priors on the unknown parameters. In the linear-link specification, the unknown parameters are $\{\bGamma,\bmu,\bLambda,\bbeta\}$; the neural-network specification additionally includes $\btheta$. The stacked coefficients are $\bLambda=(\blambda_1^T,\ldots,\blambda_G^T)^T$ of dimension $G\times(P+1)$ for linear links and $G\times(K+1)$ for neural-network links and $\bbeta=(\bbeta_1^T,\ldots,\bbeta_M^T)^T$ of dimension $M\times(P+1)$ or $M\times(K+1)$, respectively. The positive integers $M$ and $K$ are selected using the procedures in Sections~\ref{specifyM} and \ref{specifyK}. The following subsections describe the priors, likelihood, and posterior sampling procedure.}

\subsection{Prior Specification}
\label{sec:prior}
To proceed with our Bayesian analysis, we put prior distributions on the parameters. We assume independent priors for $\bGamma, \bmu, \bLambda, \bbeta,\btheta$. 
They are described below in detail.

{
\begin{enumerate}
     \item For the inclusion indicators, $\gamma_{m,g}\mid p_g\sim\operatorname{Bernoulli}(p_g)$ independently, with $p_g\sim\operatorname{Beta}(c,d)$.
     \item The component locations and all entries of the cure- and weight-coefficient matrices have independent normal priors: $\mu_m\sim\Normal(0,\sigma_\mu^2)$, $\lambda_{g,q}\sim\Normal(0,\sigma_\lambda^2)$, and $\beta_{m,q}\sim\Normal(0,\sigma_\beta^2)$, where $q$ indexes the corresponding linear predictor or neural-network output coefficient. Their variance hyperparameters independently follow $IG(a,b)$ priors.
     \item In the neural-network specification, each entry of $\btheta$ has a $\Normal(0,\sigma_\theta^2)$ prior, with $\sigma_\theta^2\sim IG(a,b)$.
\end{enumerate}
}

{We set $c=d=1$, which is a uniform prior on the treatment-specific component inclusion probability $p_g$ and gives each $\gamma_{m,g}$ a prior inclusion probability of $1/2$. We set $a=b=0.1$, giving proper, weakly-informative inverse-Gamma priors for the variance hyperparameters. As a limited sensitivity check, we repeated the analysis with $a=b=0.01$ and observed no appreciable change in the RMST estimates. Updating $\sigma_m$ would add flexibility, but preliminary experiments showed little benefit relative to the added computation. }

\subsection{Likelihood Function} 
{Let $t_i$ be the observed event or censoring time, let $g_i$ be the treatment received by subject $i$, and let $\delta_i=1$ denote censoring under our convention. Each subject contributes only through the observed treatment arm. The likelihood is 
\(
L(\bmu,\bbeta,\bGamma,\bLambda,\btheta\mid\bt,\bx)
=
\prod_{i=1}^{N}
S_{g_i}(t_i\mid\bx_i)^{\delta_i}
f_{g_i}(t_i\mid\bx_i)^{1-\delta_i},
\)
where
\(
S_g(t\mid\bx_i)
=c_g(\bx_i)+\{1-c_g(\bx_i)\}
\sum_{m=1}^{M}\pi_{i,m,g}
\left[1-\Phi\left\{\frac{\log(t)-\mu_m}{\sigma_m}\right\}\right]
\)
and
\(
f_g(t\mid\bx_i)
=\{1-c_g(\bx_i)\}
\sum_{m=1}^{M}\pi_{i,m,g}
\frac{1}{t\sigma_m}
\phi\left\{\frac{\log(t)-\mu_m}{\sigma_m}\right\}.
\)
Thus, no subject contributes likelihood factors for unobserved treatment arms. The likelihood and gradient derivations in Supplementary Sections S1 and S2 use $g_i$ explicitly.}


\subsection{Posterior Sampling}
{We use the Metropolis-adjusted Langevin algorithm (MALA) for posterior sampling by utilizing the gradients of the posterior log-likelihood for all continuous-valued parameters and update the binary indicators in $\bGamma$ element-wise from their full conditionals.}
For each simulation replicate and the AALL0434 analysis, the sampler is run for 15,000 iterations. The first 5,000 iterations are discarded as burn-in, and every tenth subsequent draw is retained, yielding 1,000 posterior draws. Proposal step sizes $\epsilon$ are tuned every 200 iterations to maintain acceptance rates between 0.45 and 0.70. Posterior convergence diagnostics are provided in Section S7 of the supplementary materials, including trace plots, parameter diagnostics, and the implementation details. 
The required gradients are provided in Supplementary Section S2, and additional sampling details are in Section S3.
 \subsection{Specification of $M$} \label{specifyM}

{For each treatment arm $g$, we apply \texttt{Mclust} to the observed log event times, which selects the component count $m_g$ by BIC. These arm-specific fits initialize a common dictionary of $M=\sum_{g=1}^G m_g$ components. Let $s_g=\sum_{r<g}m_r$ and $\I_g={s_g+1,\ldots,s_g+m_g}$. Initially, arm $g$ activates only the components in $\I_g$, giving $\bGamma$ a block-diagonal structure. During the first 1,000 iterations, only entries outside these initial blocks are updated; all entries are updated thereafter.

The component locations are initialized using their \texttt{kmeans} centroids. For a component containing $n_m>1$ observations with within-cluster sum of squares $WSS_m$, we set $\sigma_m=\sqrt{WSS_m/(n_m-1)}$; singleton or zero-scale components are assigned $\sigma_m=1$. The scales remain fixed during MCMC, while the locations, covariate-dependent weights, and inclusion indicators are updated. Preliminary experiments allowing $\sigma_m$ to update showed little improvement relative to the additional computation.

Because $M$ combines all arm-specific component counts, it provides a conservative upper dictionary. Every treatment arm retains inactive components in all simulations and the AALL0434 analysis, indicating that this upper bound is not binding. }

\subsection{Selection of $K$ in the Neural-Net Link Case}
\label{specifyK}

{We select $K$ using a five-fold predictive-log-likelihood procedure, applying the same approach in both the simulations and AALL0434 analysis. For each candidate $K$, the model is fitted to four folds and evaluated on the remaining fold. For posterior draw $b$ and held-out fold $V_v$, the criterion is
$
\ell_{K,v}^{(b)}
=
\sum_{i\in V_v}
\left[
\delta_i\log S_{g_i}^{(b)}(t_i\mid\bx_i)
+(1-\delta_i)\log f_{g_i}^{(b)}(t_i\mid\bx_i)
\right].
$

We average over posterior draws and folds and select the maximizing $K$. In the simulations, the five-fold cross-validation is performed using only the training sample; the separate 33\% test sample is reserved for final MSE evaluation.
If a boundary value is selected, we expand the grid in that direction and repeat the procedure. The resulting simulation grid is $K\in{4,6,8,13,18}$. For AALL0434, the grid is $K\in{6,8,10,15,20}$, from which $K=10$ is selected.}

\subsection{{Treatment-effect estimands and MBLP inference}} \label{sec:blp}



{Our primary estimand is the covariate-conditional RMST contrast. For treatment arm $g$,
\(
\operatorname{RMST}_g(h\mid\bx)
=
\mathbb{E}\{\min(T^{(g)},h)\mid\bX=\bx\}
=
\int_0^h S_g(t\mid\bx)\,dt,
\)
and
\(
\zeta_{gg'}^{(h)}(\bx)
=
\operatorname{RMST}_g(h\mid\bx)
-
\operatorname{RMST}_{g'}(h\mid\bx).
\)
We prespecify $h=25$ years for Simulation Setting~1, $h=5$ years for Simulation Setting~2, and $h=10$ years for AALL0434; the simulation horizons coincide with their administrative follow-up limits.

Under the proposed model,
\(
\operatorname{RMST}_g(h\mid\bx)
=
c_g(\bx)h
+\{1-c_g(\bx)\}
\sum_{m=1}^{M}\pi_{m,g}(\bx)R_m(h),
\)
where
\(
R_m(h)
=
\exp\!\left(\mu_m+\frac{\sigma_m^2}{2}\right)
\Phi\!\left\{\frac{\log(h)-\mu_m-\sigma_m^2}{\sigma_m}\right\}
+h\left[1-\Phi\!\left\{\frac{\log(h)-\mu_m}{\sigma_m}\right\}\right].
\)

We also report treatment effects on survival probability in the Supplementary Materials based on
\(
S_g(h\mid\bx)
=
c_g(\bx)
+\{1-c_g(\bx)\}
\sum_{m=1}^{M}\pi_{m,g}(\bx)
\left[1-\Phi\!\left\{\frac{\log(h)-\mu_m}{\sigma_m}\right\}\right].
\)
The survival-probability contrast is $\tau_{gg'}^{(h)}(\bx)=S_g(h\mid\bx)-S_{g'}(h\mid\bx)$. }

\ul{Marginal best linear projection:} Building on \cite{cui2023estimating}, \cite{chang2025} extended the best linear projections (BLP) approach by incorporating a thresholding step within a Bayesian framework. In this paper, a data-partitioning strategy is further added for assessing robustness in addition to applying thresholded marginal BLPs (MBLPs) marginally for each predictor.
Then, the MBLP for predictor $p$ is
$\left\{\beta_{0,p}^*, \beta_p^*\right\}=\operatorname{argmin}_{\beta_{0,p}, \beta_{p}} \mathbb{E}\left[\left(\zeta\left(\bx_i\right)-\beta_{0,p}-x_{i,p} \beta_p\right)^2\right].$



For the $b$-$th$ posterior sample, we obtain ${\zeta}_{gg', b}\left(\bx_i\right)$ based on the posterior samples of the survival probability or RMST for the $i$-$th$ patient. We next perform separate univariate regressions for ${\bzeta}_{gg', b}\left(\bx_{i}\right)$ on $x_{i,p}$, providing the estimated MBLP coefficients for the $p$-$th$ predictor $\beta_p^*$. This procedure is repeated for each subgroup, obtained by weighted k-means clustering applying the \texttt{ewkm} function from the \texttt{wskm} package \citep{wskm2014hz}.

{We summarize the MBLP coefficients using the threshold-indexed importance
measure of \citet{chang2025}. Specifically, for predictor $p$ and threshold $t$,
\(
s_{p,t}=\frac{1}{B}\sum_{b=1}^{B}\bone\{|\beta_p^{(b)}|>t\}
\)
is the posterior probability that the absolute MBLP coefficient exceeds $t$. This is an importance summary rather than a hard selection rule. At a common $t$, a larger $s_{p,t}$ indicates greater relative importance, while slower decay across $t$ indicates that the posterior coefficient remains above a wider range of effect-size thresholds.}
We demonstrate the effectiveness of this approach in Section \ref{sec:sim2}.

\section{Simulation}  \label{sec:simu}

{We evaluate covariate-conditional RMST contrasts on an independently held-out test sample. For treatment pair $(g,g')$, the criterion is
\(
\operatorname{MSE}_{gg',r}
=
\frac{1}{|T_r|}\sum_{i\in T_r}
\left[
\widehat\zeta_{gg'}^{(h)}(\bx_i)
-\zeta_{gg'}^{(h)}(\bx_i)
\right]^2,
\) for replication $r$
where $\widehat\zeta_{gg'}^{(h)}$ is the estimated contrast and $\zeta_{gg'}^{(h)}$ is its true value. We report the median of this integrated conditional CATE prediction error across 150 independently generated datasets.  

Methods that estimate each treatment contrast separately do not enforce compatibility among the resulting estimates. Specifically, Section~S8 of the Supplementary Materials illustrates this issue for pairwise causal survival forests fitted using \texttt{grf}. A binary causal survival forest is suitable for a prespecified pairwise contrast, but three separate fits do not enforce the compatibility condition $\widehat\zeta_{32}(\bx)=\widehat\zeta_{31}(\bx)-\widehat\zeta_{21}(\bx)$. Section S8 of the Supplementary Materials provides detailed diagnostics. 
Nevertheless, the proposed method yields lower RMST-contrast MSE in most comparisons while ensuring mutually coherent multi-treatment contrasts. Given the pairwise fitting-context dependence documented in Section S8, Part B, we present the \texttt{grf} results as a diagnostic benchmark rather than a like-for-like comparison.

We mainly compare our method with \texttt{flexsurvcure} \citep{flexsurvcureR}, fitting a log-normal AFT mixture-cure S-learner with treatment indicators and patient covariates. For each test subject, treatment-specific RMSTs are computed by varying the treatment indicator while holding covariates fixed, and pairwise differences are then obtained.

Section S9 of the Supplementary Materials shows results on comparing the proposed sharing mechanism, in which $\bGamma$ is updated, with a model in which $\bGamma$ remains fixed at its block-diagonal form. The latter uses disjoint component sets and gives T-learning-type estimates. Updating $\bGamma$ lowers MSE for all three contrasts under both cure-rate settings. It also lowers posterior RMST variance for almost all the arm--cure-rate comparisons except one with comparable numbers. Accordingly, all results reported for the proposed method are based on posterior updating of $\bGamma$.
}

\subsection{{Setting 1: Well-specified case}}

We generate survival times from a log-normal mixture cure model, consistent with the fitted model. 
We use sample size $n=180$, three covariates, and three equal treatment {arms}. Data are split into 67\% training and 33\% testing sets. 
We study high and low cure rate scenarios by adding offsets to the cure probability. 
{We consider $G=3$ treatment arms, so that three mutually compatible pairwise contrasts can be studied, and $P=3$ covariates to generate covariate-dependent effects. We construct nine mixture components by applying $k$-means clustering to the AALL0434 disease-free survival times and set $(\mu_m,\sigma_m)$ using the resulting clusters. The indicators $\gamma_{m,g}$ determine the active components in each arm, so that each arm uses a subset of the common dictionary. Covariates are sampled independently from a uniform distribution, and the coefficients $\beta_{m,p}$ induce covariate-dependent weights among the active components. The scenario-specific parameters are generated once and held fixed across all 150 replications: $\gamma_{m,g}\sim\Bernoulli(0.5)$, $\lambda_{g,p}\sim\Unif(-2,0)$, and $\beta_{m,p}\sim\Normal(0,1)$. Additional details are provided in Section~S5 of the Supplementary Materials.}

{Table~\ref{tab:rmst_combined} shows that, in this well-specified setting, the proposed linear and neural-network specifications have lower median RMST-contrast MSE than \texttt{flexsurvcure} for all reported contrasts and cure-rate settings. Cross-validation for selecting $K$ performs robustly and often achieves the lowest median MSE.}

\subsection{{Setting 2: Misspecified case}} \label{sec:sim2}

In this setting, survival times are generated to mimic the AALL0434 clinical trial data. 
We use the observed covariates (age, WBC, sex, CNS status), but survival times depend only on the first three predictors. 

We first partition the AALL0434 covariate space to obtain nine heterogeneous, covariate-dependent log-normal AFT cure models applying the \texttt{flexsurvcure} package. Then, for each of the three simulated treatment arms, outcomes are generated from an arm-specific, covariate-dependent mixture of these nine expert models. The partitions are calibration subsets, not treatment clusters in the fitted method. Additional details are in Section~S5 of the Supplementary Materials. The resulting Kaplan--Meier curves closely resemble those from AALL0434. The cure probability $c_g(\bx)$ is generated as in the preceding subsection. Censoring times follow an exponential distribution with rate $0.05$, and maximum follow-up is $h=5$.
Results in Table~\ref{tab:rmst_combined} again show that our method outperforms {\tt flexsurvcure}, with consistently lower median MSEs. The nonlinear link generally improves accuracy over the linear link. Similar to the well-specified case, cross-validation for selecting $K$ achieves the lowest median MSE under high cure rates, followed by $K=4$ or $K=6$. Our method for selecting $K$ demonstrates its ability to be both competitive and adaptive.

Finally, since the outcome depends only on the first three predictors, we apply the thresholded MBLP procedure from Section~\ref{sec:blp}. Table S2 in the Supplementary Materials reports the $s_{p,t}$ values for $N=961$ in the high-cure setting. As expected, CNS status shows the fastest decline in \(s_{p,t}\) as the threshold increases, consistent with its exclusion from the outcome-generating mechanism and illustrating the intended interpretation of \(s_{p,t}\). To assess robustness, we also partition the subjects into four clusters and compute the MBLPs separately within each cluster. Supplementary Tables~S3 and S4 show broadly consistent predictor-importance patterns across the clusters for the two cure settings, respectively.

\begin{table}[!htpb]
\flushleft
\small
\caption{{Median test-set MSE for the three RMST contrasts over 150 replications with $\lambda_{g,p}\sim\Unif(-2,0)$. The first panel is the model-recovery setting, and the second is the AALL0434-calibrated mixture-of-experts setting. The RMST horizons are 25 and 5 years, respectively. Here, {\tt grf} fits are based on pairwise contrasts.}}
\centering
\setlength{\tabcolsep}{3pt}
\begin{tabular}{rcccccc}
\hline
& \multicolumn{3}{c}{\textbf{High Cure Rate}} 
& \multicolumn{3}{c}{\textbf{Low Cure Rate}} \\
& $\zeta_{21}$ & $\zeta_{31}$ & $\zeta_{32}$ 
& $\zeta_{21}$ & $\zeta_{31}$ & $\zeta_{32}$ \\
\hline
\multicolumn{7}{l}{\textbf{{Setting 1: Well-specified}}} \\
\hline
\textbf{MC Linear} & 2.72 & 0.74 & 2.68 & 15.45 & 1.58 & 14.98 \\
\textbf{MC Nonlinear $K=4$} & 2.68 & 0.72 & 2.55 & 15.42 & 1.32 & 15.40 \\
\textbf{MC Nonlinear $K=6$} & 2.64 & 0.70 & 2.67 & 15.44 & 1.43 & 15.45 \\
\textbf{MC Nonlinear $K=8$} & 2.68 & 0.82 & 2.59 & 15.63 & 1.45 & 15.23 \\
\textbf{MC Nonlinear $K=13$} & 2.67 & 0.80 & 2.56 & 15.71 & 1.63 & 15.41 \\
\textbf{MC Nonlinear $K=18$} & 2.72 & 0.75 & 2.62 & 15.80 & 1.60 & 15.43 \\
\textbf{MC Nonlinear $K=K_{CV}$} & 2.61 & 0.71 & 2.55 & 15.54 & 1.37 & 14.89 \\
\hline
\textbf{{\tt grf}} & 7.71 & 1.44 & 6.93 & 47.05 & 3.15 & 42.77 \\
\textbf{{\tt flexsurvcure}} & 6.86 & 2.52 & 5.75 & 35.00 & 4.24 & 35.84 \\
\hline
\multicolumn{7}{l}{\textbf{{Setting 2: Misspecified}}} \\
\hline
\textbf{MC Linear} & 7.61 & 9.65 & 7.77 & 8.92 & 10.95 & 11.53 \\
\textbf{MC Nonlinear $K=4$} & 7.20 & 9.68 & 7.57 & 9.96 & 11.09 & 12.93 \\
\textbf{MC Nonlinear $K=6$} & 7.26 & 9.84 & 8.50 & 9.35 & 10.54 & 12.53 \\
\textbf{MC Nonlinear $K=8$} & 7.61 & 9.43 & 8.60 & 10.64 & 11.06 & 13.09 \\
\textbf{MC Nonlinear $K=13$} & 9.14 & 10.45 & 9.30 & 12.45 & 12.34 & 13.63 \\
\textbf{MC Nonlinear $K=18$} & 9.81 & 11.46 & 10.88 & 13.20 & 13.46 & 16.21 \\
\textbf{MC Nonlinear $K=K_{CV}$} & 7.06 & 8.64 & 7.72 & 9.68 & 10.32 & 12.21 \\
\hline
\textbf{{\tt grf}} & 6.24 & 10.36 & 7.29 & 12.59 & 13.66 & 15.43 \\
\textbf{{\tt flexsurvcure}} & 9.59 & 11.28 & 12.46 & 20.23 & 18.33 & 23.99 \\
\hline
\end{tabular}
\label{tab:rmst_combined}
\end{table}

\section{AALL0434 Data Analysis} \label{sec:real}

Following \cite{winter2018improved}, we include age, sex, white blood cell count, CNS status, and risk status as covariates in the analysis.
Before model fitting, continuous covariates are rescaled to \([0,1]\) using min-max scaling, while categorical covariates are represented using one-hot encoding.
{We fit both the linear and neural-network specifications. The five-fold predictive-log-likelihood procedure in Section~\ref{specifyK} selected the neural-network model with $K=10$, which is used for the results below.}

The treatment arms, C-MTX, C-MTX combined with nelarabine, HDMTX, and HDMTX combined with nelarabine, are referred to as Treatments 1 through 4, respectively. 
{We summarize posterior inference in three complementary ways: covariate-conditional contrasts in 10-year RMST, treatment-specific survival curves across selected covariate strata, and marginal best linear projection (MBLP) summaries. Following Section~\ref{sec:blp}, we compute $s_{p,t}$ for three prespecified contrasts, $\widehat{\bzeta}_{31}$, $\widehat{\bzeta}_{21}$, and $\widehat{\bzeta}_{43}$, using the full data and two partitions. The first compares HDMTX with C-MTX without nelarabine, and the other two compare the addition of nelarabine within each methotrexate regimen.}



\begin{figure}[!htpb]
\centering

\begin{subfigure}[t]{0.32\textwidth}
\includegraphics[width=\textwidth, trim=0cm 0cm 1cm 1.55cm, clip=true]{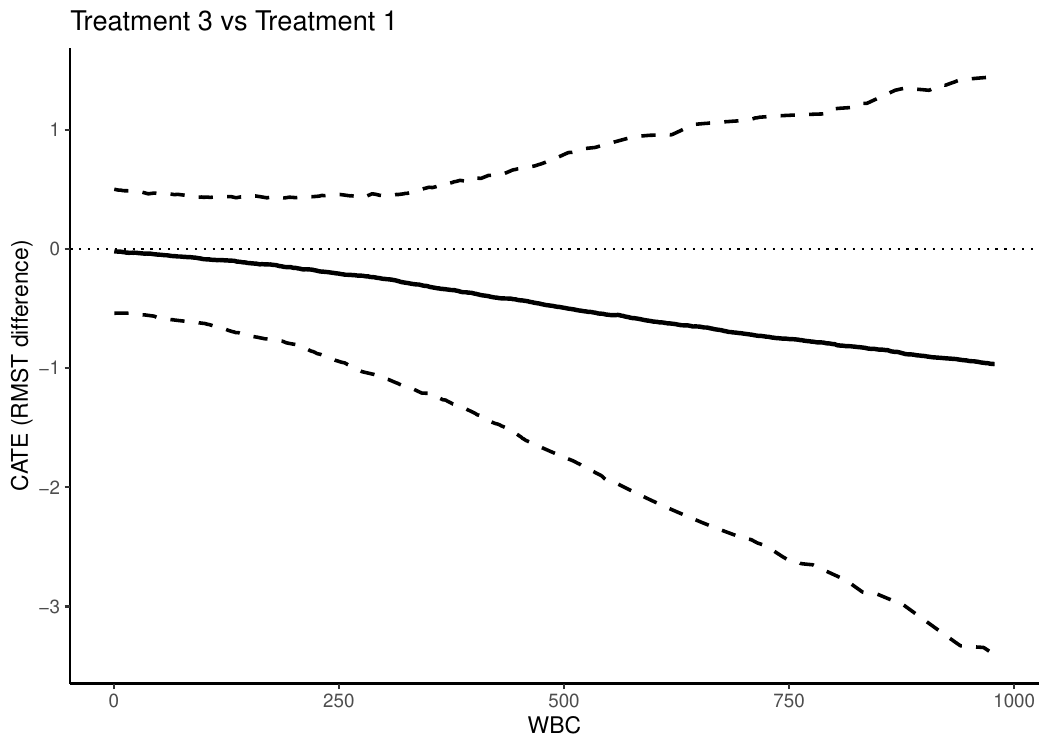}
\caption{$\widehat{\bzeta}_{31}$: HDMTX vs C-MTX (WBC)}
\label{fig:cate_wbc31_comb}
\end{subfigure}
\hfill
\begin{subfigure}[t]{0.32\textwidth}
\includegraphics[width=\textwidth, trim=0cm 0cm 1cm 1.55cm, clip=true]{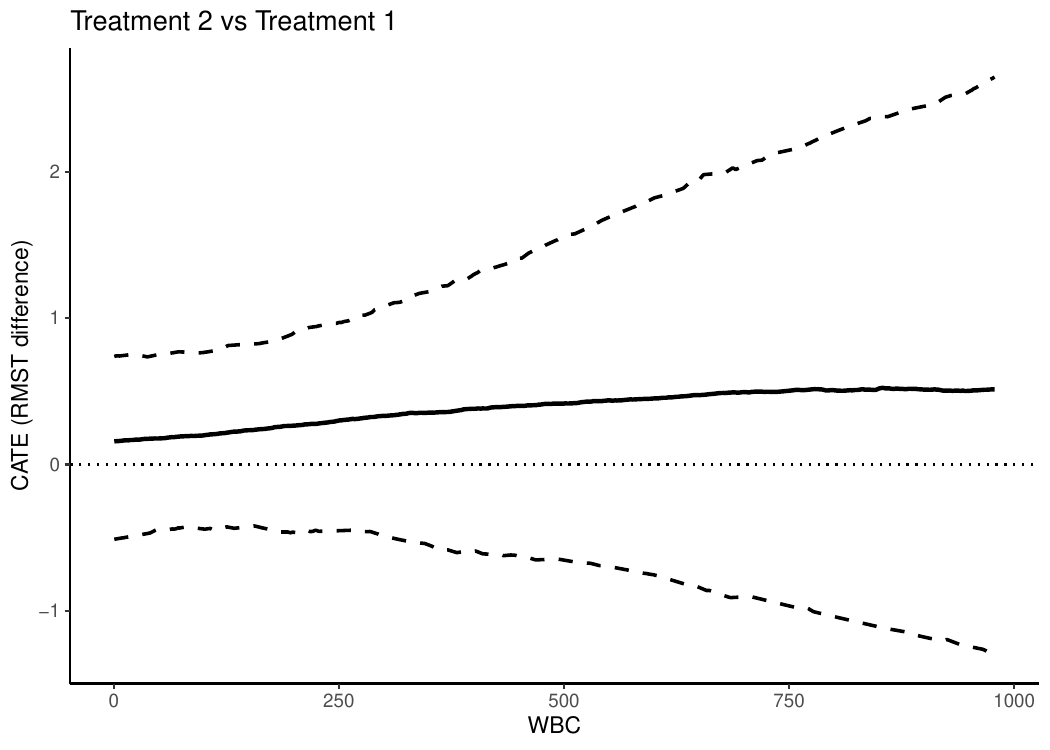}
\caption{$\widehat{\bzeta}_{21}$: C-MTX w/ nelarabine vs C-MTX (WBC)}
\label{fig:cate_wbc21_comb}
\end{subfigure}
\hfill
\begin{subfigure}[t]{0.32\textwidth}
\includegraphics[width=\textwidth, trim=0cm 0cm 1cm 1.55cm, clip=true]{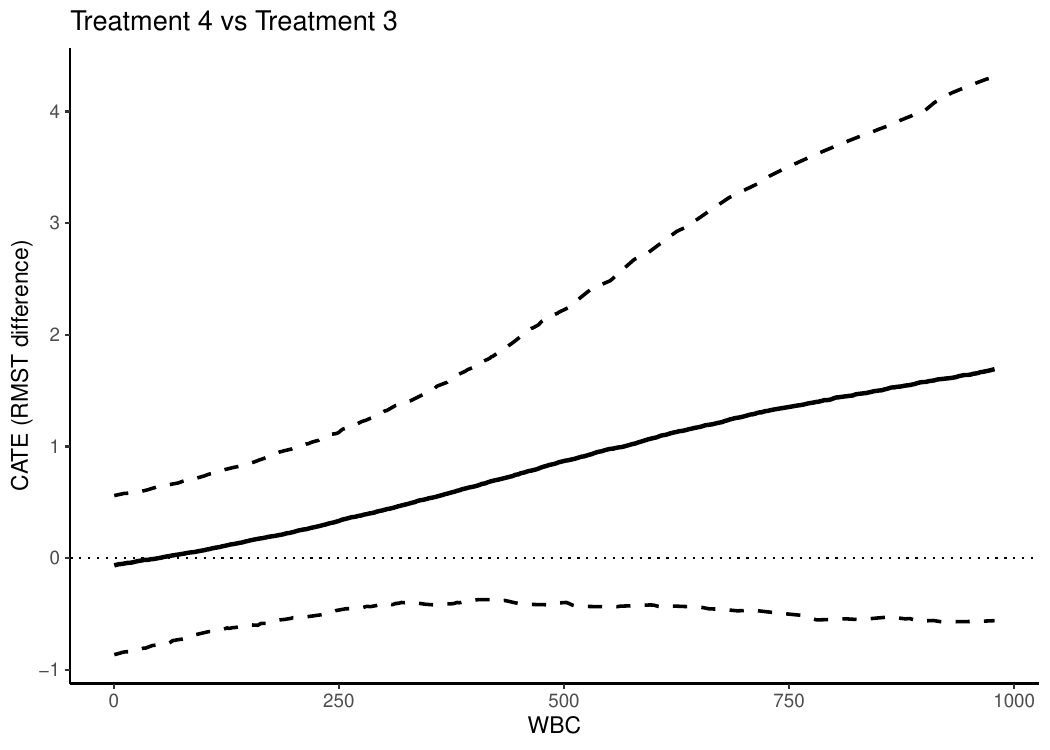}
\caption{$\widehat{\bzeta}_{43}$: HDMTX w/ nelarabine vs HDMTX (WBC)}
\label{fig:cate_wbc43_comb}
\end{subfigure}

\vskip\baselineskip

\begin{subfigure}[t]{0.32\textwidth}
\includegraphics[width=\textwidth, trim=0cm 0cm 1cm 1.55cm, clip=true]{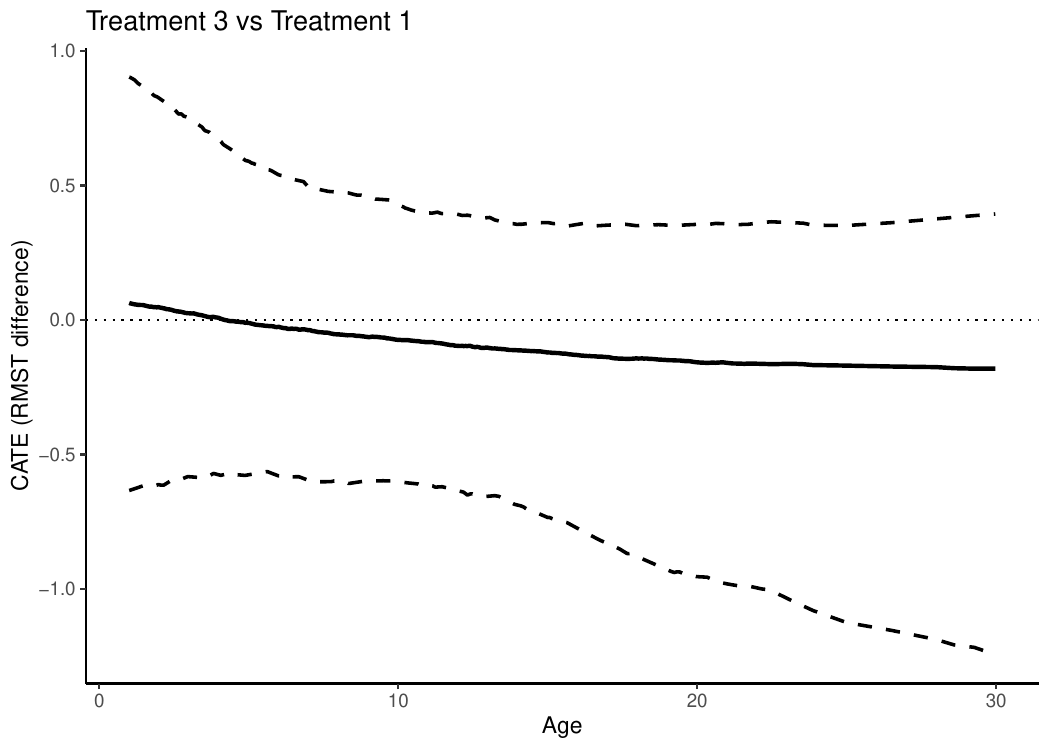}
\caption{$\widehat{\bzeta}_{31}$: HDMTX vs C-MTX (Age)}
\label{fig:cate_age31_comb}
\end{subfigure}
\hfill
\begin{subfigure}[t]{0.32\textwidth}
\includegraphics[width=\textwidth, trim=0cm 0cm 1cm 1.55cm, clip=true]{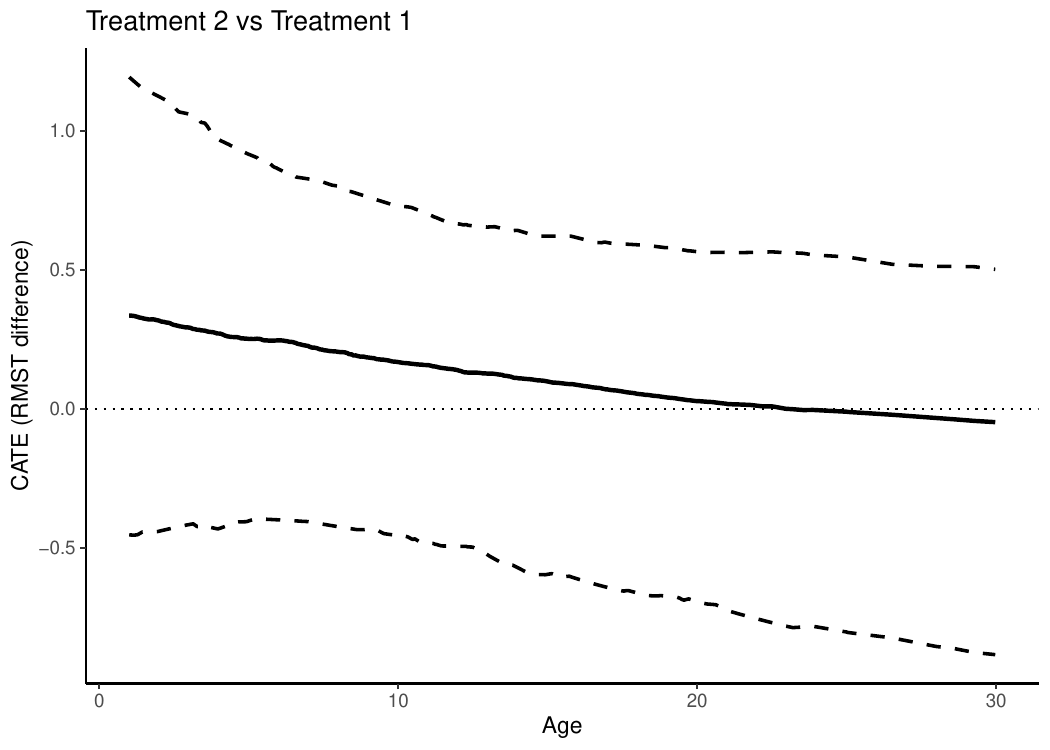}
\caption{$\widehat{\bzeta}_{21}$: C-MTX w/ nelarabine vs C-MTX (Age)}
\label{fig:cate_age21_comb}
\end{subfigure}
\hfill
\begin{subfigure}[t]{0.32\textwidth}
\includegraphics[width=\textwidth, trim=0cm 0cm 1cm 1.55cm, clip=true]{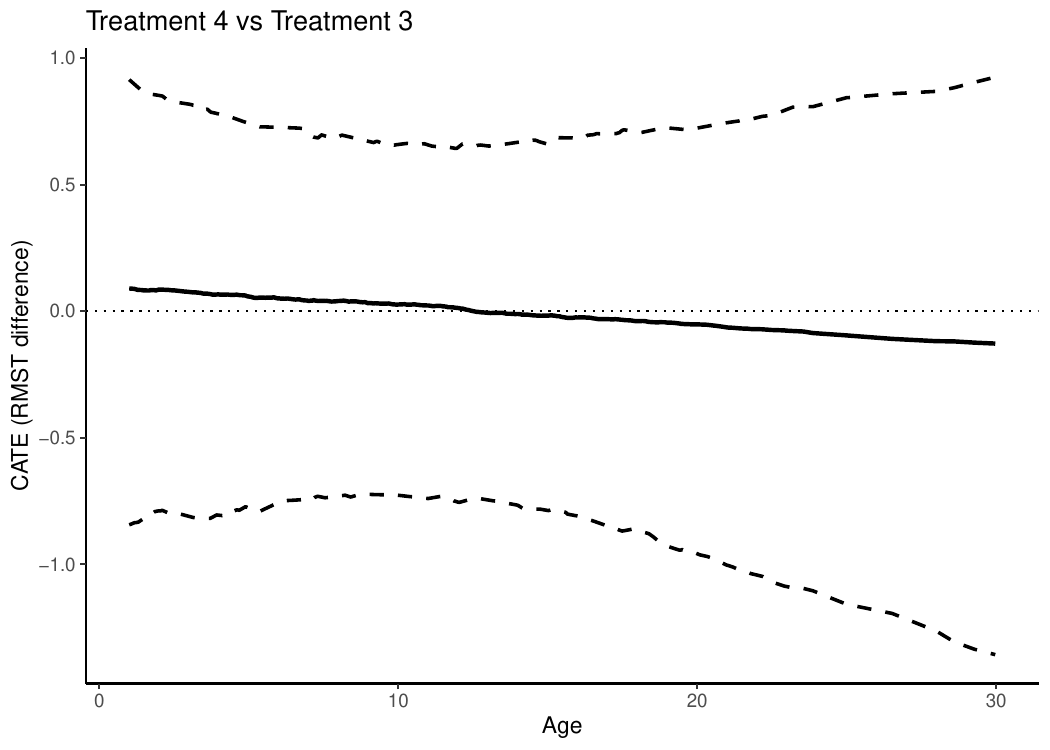}
\caption{$\widehat{\bzeta}_{43}$: HDMTX w/ nelarabine vs HDMTX (Age)}
\label{fig:cate_age43_comb}
\end{subfigure}

\caption{
{Posterior mean covariate-conditional treatment effects (solid lines) with 95\% posterior credible bands (dashed lines) for 10-year RMST, computed from 1,000 retained posterior draws. Treatments 1--4 are C-MTX, C-MTX with nelarabine, HDMTX, and HDMTX with nelarabine, respectively. In the top row, WBC varies and age is fixed at its sample median of 10 years. In the bottom row, age varies and WBC is fixed at its sample median of $50{,}000/\mu\mathrm{L}$. In every panel, sex, CNS status, and risk group are fixed at their observed modes: male, CNS1, and intermediate risk. The analysis uses $N=1{,}022$ patients and neural-network covariate links with $K=10$.}
}
\label{fig:cate_combined}
\end{figure}

{\ul{CATE (RMST in years) comparison:}} 
{We plot posterior mean CATEs and 95\% credible bands against WBC count and age separately, with the remaining covariates fixed as specified in Figure~\ref{fig:cate_combined}. Thus, each panel varies one continuous covariate at a time.

In the top row, the posterior mean RMST contrast for HDMTX versus C-MTX decreases as WBC increases, whereas the contrast for adding nelarabine to HDMTX increases. The contrast for adding nelarabine to C-MTX is comparatively stable across WBC. These patterns show that the estimated association between WBC and the RMST contrast differs across treatment comparisons, with uncertainty represented by the credible bands.}

{In the bottom row, the posterior mean contrasts for HDMTX versus C-MTX and for adding nelarabine to C-MTX decrease with age before becoming relatively flat at older ages. The contrast for adding nelarabine to HDMTX remains close to zero across most of the age range. Thus, the fitted age-related variation is more pronounced for the first two comparisons.}

\begin{figure}[!htpb]
\centering

\begin{subfigure}[t]{0.5\textwidth}
\centering
\includegraphics[width=\textwidth, trim=0cm 0cm 0cm 1.3cm, clip=true]{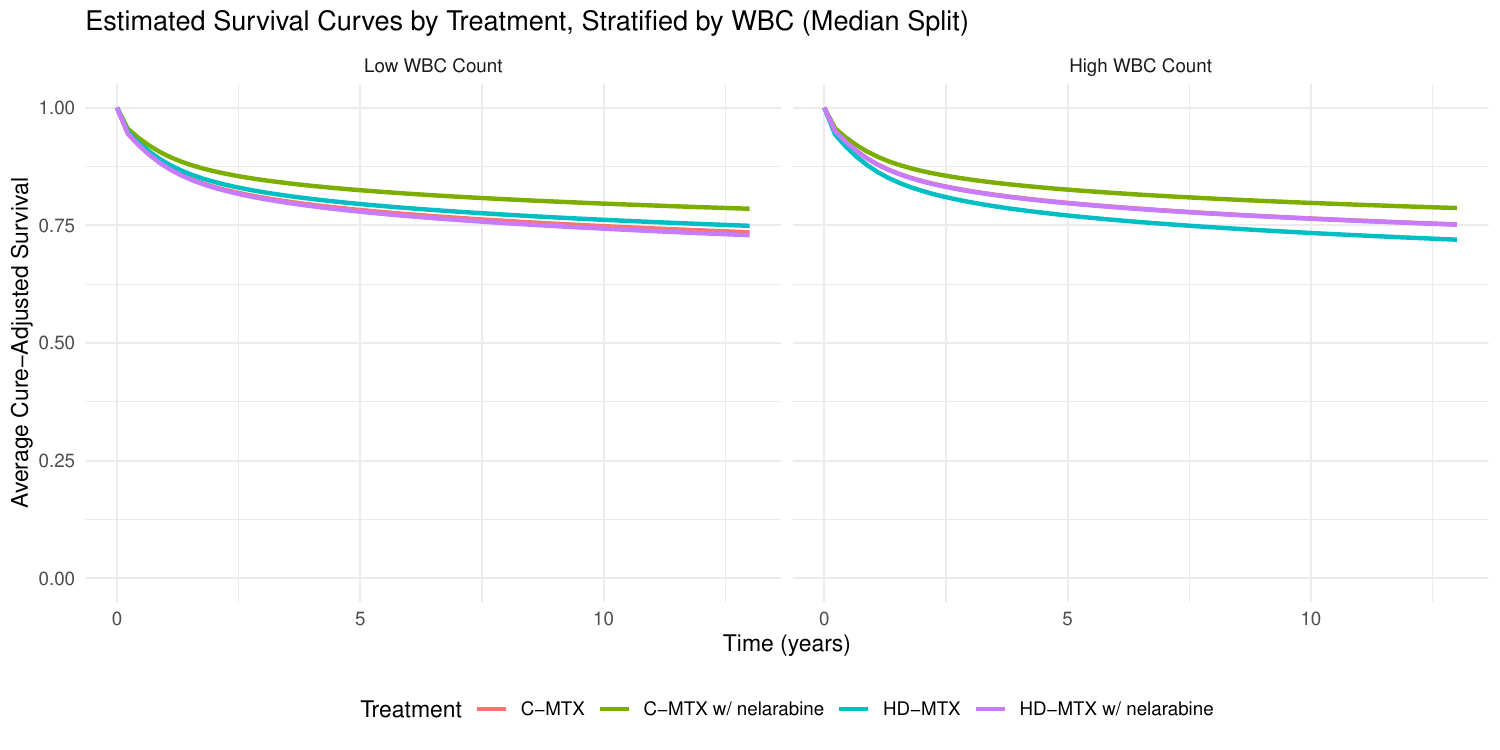}
\caption{Stratified by WBC}
\label{fig:km_wbc_comb}
\end{subfigure}
\hfill
\begin{subfigure}[t]{0.5\textwidth}
\centering
\includegraphics[width=\textwidth, trim=0cm 0cm 0cm 1.3cm, clip=true]{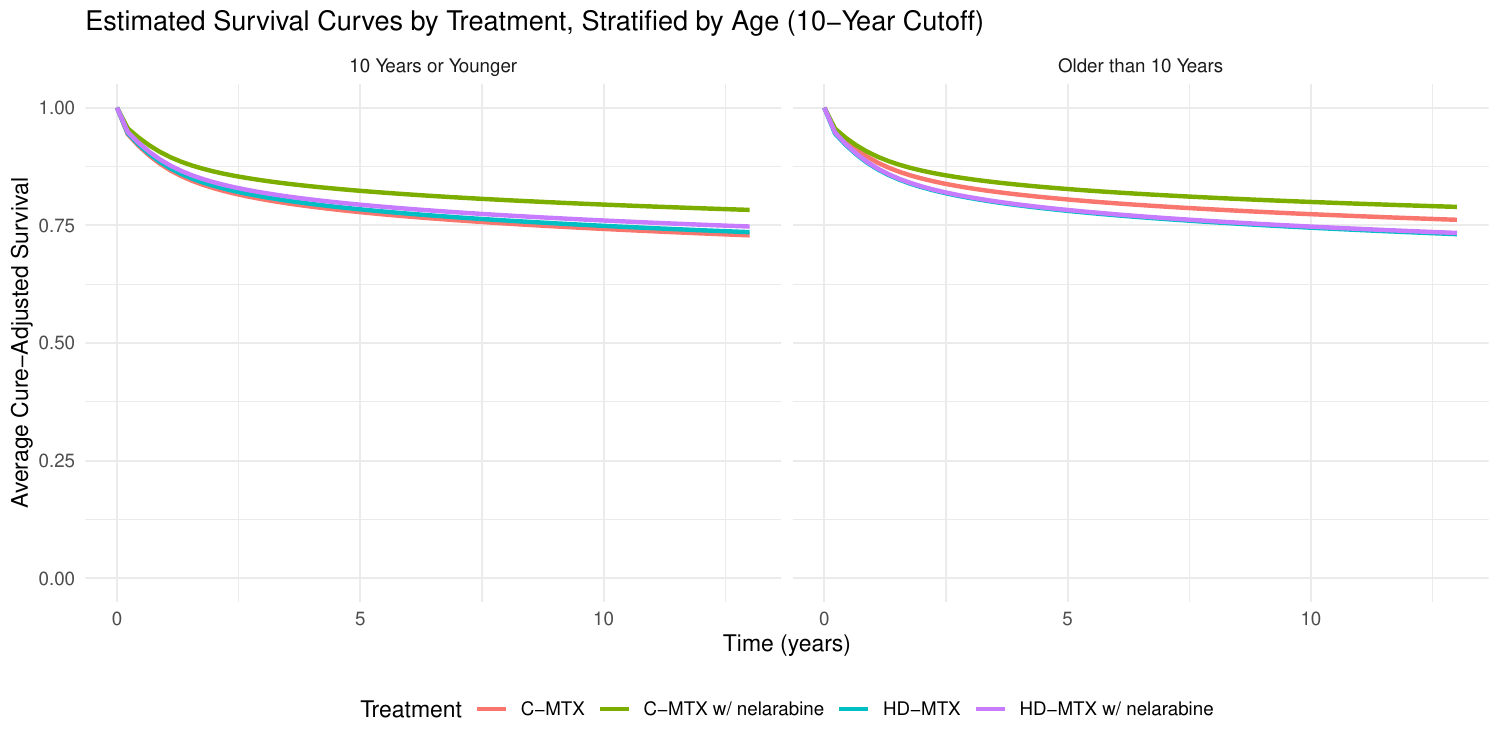}
\caption{Stratified by Age}
\label{fig:km_age_comb}
\end{subfigure}
\hfill
\begin{subfigure}[t]{0.5\textwidth}
\centering
\includegraphics[width=\textwidth, trim=0cm 0cm 0cm 1.3cm, clip=true]{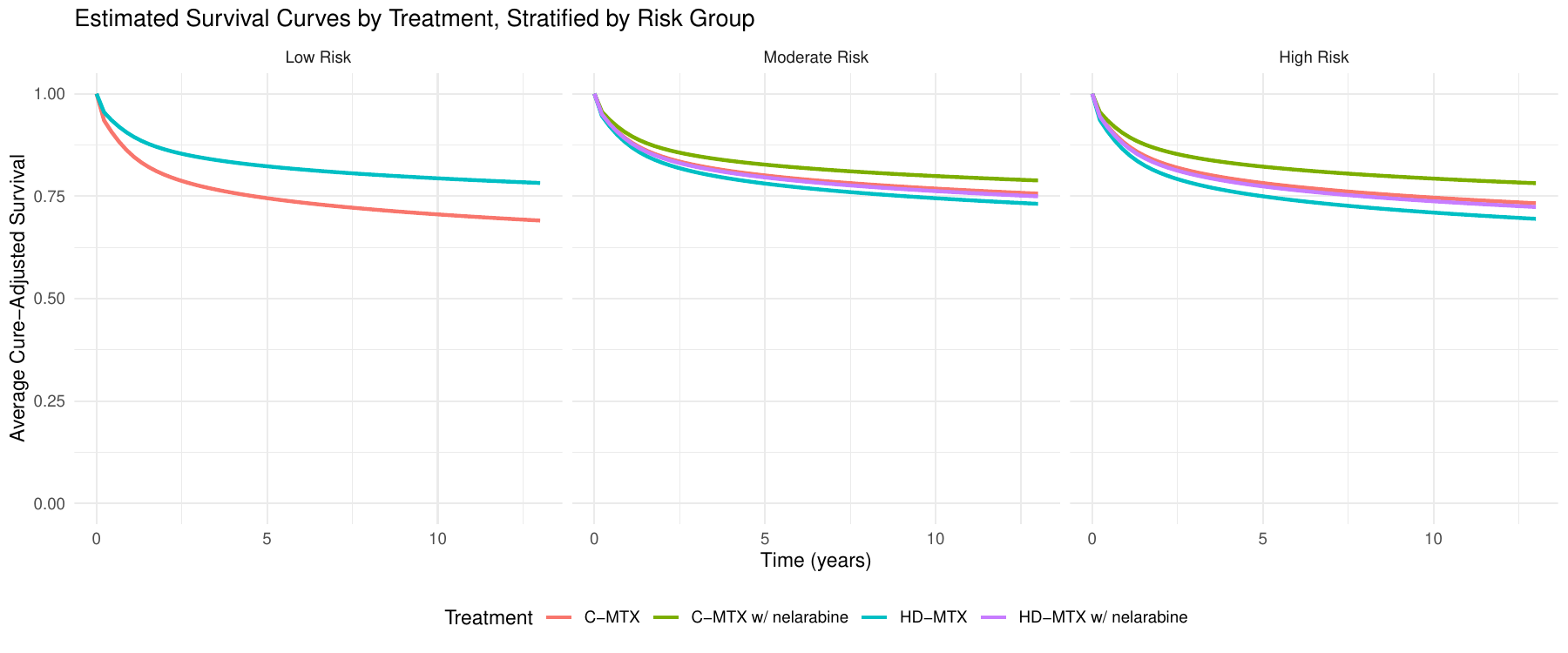}
\caption{Stratified by Risk Group}
\label{fig:km_risk_comb}
\end{subfigure}
\caption{
Estimated treatment-specific average cure-adjusted survival curves under different methotrexate regimens with/without nelarabine from the AALL0434 clinical trial ({$N=1{,}022$}). 
Survival probabilities are obtained from our proposed {treatment-specific} mixture cure model with neural network links ($K=10$), using {1,000 retained posterior draws} and evaluated on equally spaced time points from $t=0$ to $t=10$ years.
\textbf{Top:} Stratification by white blood cell (WBC) count, where high WBC is defined as $\geq 50{,}000\,\mu$L (sample median).  
\textbf{Middle:} Stratification by age (older defined as $>10$ years).  
\textbf{Bottom:} Stratification by risk group (low, moderate, high).
{Differences between curves summarize posterior variation in the fitted marginal survival distributions across treatment arms and covariate strata.}
}
\label{fig:km_combined}
\end{figure}

{\ul{Estimated survival curves comparison:}} 
{We next average the subject-specific posterior survival probabilities within the WBC, age, and risk strata shown in Figure~\ref{fig:km_combined}. Figure~\ref{fig:km_combined}(a) shows greater separation among the estimated treatment-specific survival curves in the high-WBC stratum. In the low-WBC stratum, the curves are more closely aligned, with the C-MTX plus nelarabine curve lying slightly above the others over much of follow-up. This agrees with the stronger variation across WBC in the CATE displays.

In Figure~\ref{fig:km_combined}(b), the nelarabine arms have higher posterior mean survival curves than their corresponding non-nelarabine arms in the younger stratum, whereas the ordering changes in the older stratum. Nevertheless, the between-arm areas remain broadly similar across the two age strata, consistent with the more modest age-related variation in some of the RMST contrasts. Figure~\ref{fig:km_combined}(c) shows that separation between the HDMTX and C-MTX curves differs across risk strata. Comparisons involving nelarabine are not shown for low-risk patients because they were not eligible for the nelarabine randomization. Thus, the stratified survival curves provide a second view of the covariate-dependent patterns seen in the RMST contrasts.}

{\ul{MBLP comparison:}} 
{Table~\ref{tab:test} reports $s_{p,t}$ for three treatment contrasts using the full dataset and both partitions; Supplementary Table~S6 reports the remaining contrasts. We use these values as comparative importance summaries rather than as a hard selection rule.} {Across the displayed thresholds, WBC generally has the largest $s_{p,t}$ for all three contrasts. Low versus moderate risk and CNS status are also prominent for $\widehat{\bzeta}_{31}$, while age is among the larger contributors for $\widehat{\bzeta}_{21}$ and $\widehat{\bzeta}_{43}$. The relative ordering and decay patterns are broadly similar across the two partitions. The prominence of WBC is consistent with its established prognostic role in pediatric leukemia \citep{liu2022bayesian}. Its prognostic role alone does not establish treatment-effect modification; here, the CATE and MBLP summaries describe its model-based association with the RMST contrasts.}
\begin{table}[!htpb]
\caption{The $s_{p,t}$-values for $\widehat{\bzeta}_{31}$, $\widehat{\bzeta}_{21}$, and $\widehat{\bzeta}_{43}$ under different choices of thresholds ($t$) and predictors for the AALL0434 dataset. We present the thresholding results for the full data and the different data partitions. Entries associated with some of the contrasts for the `Low Risk' column are omitted because low-risk patients were not eligible for
nelarabine. 
}
\flushleft
\centering
\scriptsize
\setlength{\tabcolsep}{3pt} 
\begin{tabular}{ccccccccccccccccccc}
\hline
& \multicolumn{6}{c}{\textbf{Full-data}} & \multicolumn{6}{c}{\textbf{Partition 1}} & \multicolumn{6}{c}{\textbf{Partition 2}} \\ 
  \hline
 & Age & \makecell{CNS \\ Status} & Sex & \makecell{WBC \\Count} & \makecell{Low \\Risk} & \makecell{High \\Risk} & Age & \makecell{CNS \\ Status} & Sex & \makecell{WBC \\Count} & \makecell{Low \\Risk} & \makecell{High \\Risk} & Age & \makecell{CNS \\ Status} & Sex & \makecell{WBC \\Count} & \makecell{Low \\Risk} & \makecell{High \\Risk} \\ 
\hline
Threshold & \multicolumn{18}{c}{$\widehat{\bzeta}_{31}$} \\ 
\hline
0.05 & 0.94 & 0.97 & 0.92 & 0.99 & 0.99 & 0.95 & 0.97 & 0.98 & 0.93 & 0.99 & 0.99 & 0.95 & 0.93 & 0.97 & 0.90 & 0.99 & 0.98 & 0.95 \\
0.15 & 0.81 & 0.93 & 0.76 & 0.97 & 0.95 & 0.83 & 0.88 & 0.95 & 0.75 & 0.97 & 0.97 & 0.84 & 0.83 & 0.89 & 0.74 & 0.97 & 0.92 & 0.83 \\
0.25 & 0.70 & 0.88 & 0.58 & 0.95 & 0.92 & 0.73 & 0.80 & 0.90 & 0.63 & 0.96 & 0.94 & 0.76 & 0.73 & 0.83 & 0.55 & 0.95 & 0.88 & 0.73 \\
0.35 & 0.58 & 0.83 & 0.44 & 0.92 & 0.86 & 0.64 & 0.71 & 0.86 & 0.50 & 0.94 & 0.91 & 0.66 & 0.62 & 0.76 & 0.40 & 0.93 & 0.82 & 0.63 \\
0.40 & 0.53 & 0.81 & 0.38 & 0.91 & 0.84 & 0.60 & 0.67 & 0.85 & 0.45 & 0.93 & 0.89 & 0.63 & 0.57 & 0.72 & 0.33 & 0.91 & 0.79 & 0.57 \\
0.50 & 0.45 & 0.74 & 0.26 & 0.88 & 0.78 & 0.49 & 0.58 & 0.79 & 0.33 & 0.91 & 0.84 & 0.54 & 0.49 & 0.63 & 0.22 & 0.89 & 0.71 & 0.49 \\
0.60 & 0.38 & 0.67 & 0.18 & 0.86 & 0.71 & 0.41 & 0.51 & 0.73 & 0.24 & 0.88 & 0.79 & 0.45 & 0.42 & 0.53 & 0.14 & 0.87 & 0.62 & 0.41 \\
0.70 & 0.32 & 0.56 & 0.12 & 0.83 & 0.62 & 0.34 & 0.46 & 0.68 & 0.19 & 0.85 & 0.74 & 0.38 & 0.36 & 0.43 & 0.08 & 0.84 & 0.52 & 0.33 \\
0.80 & 0.26 & 0.48 & 0.07 & 0.80 & 0.54 & 0.27 & 0.41 & 0.60 & 0.14 & 0.83 & 0.67 & 0.32 & 0.29 & 0.36 & 0.05 & 0.82 & 0.44 & 0.26 \\
0.90 & 0.21 & 0.41 & 0.05 & 0.76 & 0.44 & 0.21 & 0.37 & 0.54 & 0.11 & 0.80 & 0.59 & 0.27 & 0.25 & 0.29 & 0.03 & 0.80 & 0.35 & 0.22 \\
1.00 & 0.17 & 0.33 & 0.04 & 0.73 & 0.36 & 0.18 & 0.32 & 0.47 & 0.07 & 0.78 & 0.51 & 0.22 & 0.21 & 0.21 & 0.02 & 0.76 & 0.27 & 0.18 \\
\hline
& \multicolumn{18}{c}{$\widehat{\bzeta}_{21}$} \\ 
\hline
0.05 & 0.93 & 0.94 & 0.94 & 0.95 & -- & 0.93 & 0.93 & 0.94 & 0.94 & 0.96 & -- & 0.94 & 0.93 & 0.93 & 0.94 & 0.96 & -- & 0.93 \\
0.15 & 0.82 & 0.81 & 0.82 & 0.88 & -- & 0.80 & 0.84 & 0.81 & 0.81 & 0.88 & -- & 0.81 & 0.81 & 0.82 & 0.84 & 0.87 & -- & 0.79 \\
0.25 & 0.70 & 0.67 & 0.70 & 0.81 & -- & 0.68 & 0.75 & 0.69 & 0.70 & 0.80 & -- & 0.71 & 0.70 & 0.68 & 0.72 & 0.79 & -- & 0.67 \\
0.35 & 0.60 & 0.56 & 0.57 & 0.71 & -- & 0.58 & 0.65 & 0.58 & 0.56 & 0.72 & -- & 0.61 & 0.59 & 0.55 & 0.60 & 0.71 & -- & 0.58 \\
0.40 & 0.56 & 0.50 & 0.52 & 0.68 & -- & 0.53 & 0.60 & 0.52 & 0.50 & 0.69 & -- & 0.57 & 0.54 & 0.49 & 0.54 & 0.68 & -- & 0.53 \\
0.50 & 0.49 & 0.38 & 0.41 & 0.61 & -- & 0.44 & 0.53 & 0.43 & 0.41 & 0.60 & -- & 0.46 & 0.45 & 0.39 & 0.44 & 0.59 & -- & 0.44 \\
0.60 & 0.41 & 0.30 & 0.31 & 0.52 & -- & 0.36 & 0.46 & 0.34 & 0.33 & 0.53 & -- & 0.38 & 0.39 & 0.30 & 0.34 & 0.53 & -- & 0.36 \\
0.70 & 0.33 & 0.24 & 0.25 & 0.45 & -- & 0.28 & 0.40 & 0.26 & 0.27 & 0.46 & -- & 0.31 & 0.33 & 0.23 & 0.26 & 0.47 & -- & 0.28 \\
0.80 & 0.28 & 0.18 & 0.18 & 0.40 & -- & 0.23 & 0.34 & 0.20 & 0.21 & 0.40 & -- & 0.26 & 0.27 & 0.17 & 0.18 & 0.41 & -- & 0.23 \\
0.90 & 0.24 & 0.13 & 0.14 & 0.34 & -- & 0.19 & 0.29 & 0.17 & 0.16 & 0.35 & -- & 0.20 & 0.22 & 0.12 & 0.13 & 0.35 & -- & 0.18 \\
1.00 & 0.20 & 0.09 & 0.09 & 0.28 & -- & 0.14 & 0.25 & 0.12 & 0.11 & 0.30 & -- & 0.17 & 0.19 & 0.08 & 0.09 & 0.30 & -- & 0.14 \\
\hline
& \multicolumn{18}{c}{$\widehat{\bzeta}_{43}$} \\ 
\hline
0.05 & 0.96 & 0.94 & 0.92 & 1.00 & -- & 0.96 & 0.96 & 0.96 & 0.93 & 1.00 & -- & 0.95 & 0.95 & 0.94 & 0.93 & 0.99 & -- & 0.96 \\
0.15 & 0.89 & 0.83 & 0.80 & 0.99 & -- & 0.84 & 0.88 & 0.86 & 0.81 & 0.99 & -- & 0.86 & 0.87 & 0.80 & 0.79 & 0.98 & -- & 0.84 \\
0.25 & 0.81 & 0.73 & 0.66 & 0.98 & -- & 0.75 & 0.82 & 0.76 & 0.67 & 0.98 & -- & 0.76 & 0.75 & 0.68 & 0.68 & 0.97 & -- & 0.75 \\
0.35 & 0.75 & 0.61 & 0.56 & 0.97 & -- & 0.64 & 0.75 & 0.68 & 0.57 & 0.97 & -- & 0.67 & 0.67 & 0.56 & 0.56 & 0.96 & -- & 0.66 \\
0.40 & 0.71 & 0.56 & 0.49 & 0.97 & -- & 0.61 & 0.72 & 0.64 & 0.52 & 0.97 & -- & 0.62 & 0.61 & 0.51 & 0.50 & 0.95 & -- & 0.61 \\
0.50 & 0.64 & 0.47 & 0.39 & 0.96 & -- & 0.51 & 0.64 & 0.56 & 0.42 & 0.96 & -- & 0.54 & 0.55 & 0.41 & 0.40 & 0.94 & -- & 0.53 \\
0.60 & 0.57 & 0.38 & 0.28 & 0.95 & -- & 0.43 & 0.58 & 0.48 & 0.34 & 0.95 & -- & 0.47 & 0.47 & 0.33 & 0.33 & 0.93 & -- & 0.43 \\
0.70 & 0.51 & 0.32 & 0.22 & 0.94 & -- & 0.35 & 0.53 & 0.41 & 0.26 & 0.94 & -- & 0.39 & 0.40 & 0.26 & 0.26 & 0.92 & -- & 0.36 \\
0.80 & 0.45 & 0.25 & 0.17 & 0.93 & -- & 0.26 & 0.46 & 0.34 & 0.21 & 0.93 & -- & 0.32 & 0.35 & 0.20 & 0.20 & 0.90 & -- & 0.26 \\
0.90 & 0.38 & 0.18 & 0.13 & 0.91 & -- & 0.21 & 0.41 & 0.29 & 0.16 & 0.92 & -- & 0.26 & 0.30 & 0.16 & 0.14 & 0.87 & -- & 0.20 \\
1.00 & 0.33 & 0.14 & 0.09 & 0.89 & -- & 0.16 & 0.37 & 0.24 & 0.11 & 0.90 & -- & 0.22 & 0.24 & 0.11 & 0.10 & 0.85 & -- & 0.15 \\
\hline
\end{tabular}
\label{tab:test}
\end{table}

\begin{table}[!htpb]
\centering
\small
\setlength{\tabcolsep}{3pt}
\caption{{Posterior mean MBLP coefficients for the 10-year RMST contrasts $\widehat{\bzeta}_{31}$, $\widehat{\bzeta}_{21}$, and $\widehat{\bzeta}_{43}$ in AALL0434. Treatment arms 1--4 are C-MTX, C-MTX with nelarabine, HDMTX, and HDMTX with nelarabine, respectively.  Entries associated with some of the contrasts for the `Low vs Moderate Risk' row are omitted because low-risk patients were not eligible for
nelarabine.}
}
\begin{tabular}{rccccccccc}
  \hline
 & \multicolumn{3}{c}{\textbf{Full-data}} & \multicolumn{3}{c}{\textbf{Partition 1}} & \multicolumn{3}{c}{\textbf{{Partition 2}}} \\ 
  \hline
  \textbf{Predictor}  &  $\widehat{\bzeta}_{31}$ & $\widehat{\bzeta}_{21}$ & $\widehat{\bzeta}_{43}$ & $\widehat{\bzeta}_{31}$ & $\widehat{\bzeta}_{21}$ & $\widehat{\bzeta}_{43}$ & $\widehat{\bzeta}_{31}$ & $\widehat{\bzeta}_{21}$ & $\widehat{\bzeta}_{43}$\\ 
  \hline
  Age & -0.86 & -1.01 & -0.50 & -1.44 & -0.81 & 0.24 & -1.24 & -0.98 & 0.10 \\ 
  CNS Status 2 vs 1 & -1.29 & -0.62 & 0.54 & -1.64 & -0.67 & 0.88 & -1.00 & -0.60 & 0.20 \\ 
  Sex & 0.48 & 0.79 & 0.46 & 0.55 & 0.71 & 0.39 & 0.42 & 0.85 & 0.57 \\ 
  WBC Count & -2.50 & 0.54 & 4.24 & -2.74 & 0.41 & 4.36 & -2.83 & 0.61 & 4.69 \\ 
  Low vs Moderate Risk & 1.21 & - & - & 1.42 & - & - & 1.09 & - & - \\
  High vs Moderate Risk & -0.47 & 0.40 & 0.32 & -0.47 & 0.30 & 0.30 & -0.50 & 0.47 & 0.39 \\ 
\end{tabular}
\label{tab:aall_mblp}
\end{table}

We further inspect the direction and magnitude of these associations using the posterior mean MBLP coefficients in Table~\ref{tab:aall_mblp}.
{The WBC coefficient is negative for $\widehat{\bzeta}_{31}$ and positive for $\widehat{\bzeta}_{21}$ and $\widehat{\bzeta}_{43}$ in the full data and both partitions. Thus, as WBC increases, the fitted HDMTX-versus-C-MTX contrast decreases, whereas the two contrasts for adding nelarabine increase. Age has negative coefficients for $\widehat{\bzeta}_{31}$ and $\widehat{\bzeta}_{21}$ throughout; its coefficient for $\widehat{\bzeta}_{43}$ is smaller and changes sign across the full-data and partition-specific summaries. For $\widehat{\bzeta}_{31}$, the positive low-versus-moderate-risk coefficient is also stable across partitions, indicating a larger fitted HDMTX-versus-C-MTX contrast for low-risk profiles. Nelarabine contrasts are not reported for low-risk profiles because those patients were ineligible for the nelarabine randomization.

Overall, the CATE curves, stratified survival curves, and MBLP summaries show broadly consistent posterior patterns involving WBC, age, and risk group, although the magnitude of some associations differs across partitions. These exploratory findings complement earlier AALL0434 analyses \citep{winter2018improved,dunsmore2020children} while focusing on covariate-dependent RMST contrasts. Clinical use would require joint profile-specific posterior predictions together with treatment eligibility, toxicity, and other clinical considerations.}

\section{Discussion} \label{sec:discuss}

{We propose a Bayesian multi-treatment mixture-cure model with adaptively shared components and linear or neural-network covariate links, while retaining treatment-specific cure probabilities and mixture weights. Joint estimation ensures that all RMST contrasts are mutually compatible. We also study the need for the adaptive sharing of mixture components over T-learning-type model fitting in the supplementary materials.  
We note that the cure-fraction estimation depends on adequate tail information and the assumed mixture-cure structure. Although AALL0434 has a long follow-up with clear tail flattening, uncertainty about the long-term cure fraction necessarily remains under finite follow-up.



Promotion-time and hierarchical or dependent Dirichlet-process cure models provide important alternatives \citep{chen1999new,ibrahim2001bayesian,yin2005cure}; however, matched comparisons would require common specifications for event-time distributions, covariate effects, cross-arm sharing, and multi-treatment contrasts.  Specifically, adapting them to support covariate-dependent cross-treatment sharing and mutually compatible RMST contrasts is an important direction for future work. It will also be important to study their relative computational or statistical performances.

The simulations evaluate integrated prediction error for covariate-conditional RMST contrasts. Marginal effects can be obtained by marginalizing $\operatorname{RMST}_g(h\mid\bX)$ over a target covariate distribution. An important future direction will be to evaluate the frequentist operating characteristics of the proposed model for these marginal estimands, including bias, coverage, and interval width.
Future work may consider mixture-of-experts extensions in which both component survival distributions and their weights vary with covariates \citep{jordan1994hierarchical,svensn2003bayesian}. Motivated by the AALL0434 application, the present study focuses on settings with a cure fraction. Evaluating the proposed framework without the cure component is an important direction for future work.}




\clearpage
\appendix
\renewcommand{\thesection}{S\arabic{section}}
\renewcommand{\thesubsection}{S\arabic{section}.\arabic{subsection}}
\setcounter{section}{0}
\setcounter{table}{0}
\setcounter{figure}{0}
\renewcommand{\thetable}{S\arabic{table}}
\renewcommand{\thefigure}{S\arabic{figure}}
\begin{center}
{\Large\bfseries Supplementary Materials}\\[0.5em]
{\large Bayesian Neural-Net-Assisted Multi-Treatment Mixture Cure Survival Model with Application in Pediatric Oncology}
\end{center}
\bigskip
\section{Likelihood derivation of the mixture cure model} \label{supp::likelihood}

Let $g_i\in\{1,\ldots,G\}$ denote the treatment received by subject $i$, and
let $\delta_i=1$ denote right censoring (so $1-\delta_i=1$ denotes an
observed event). For component $m$, define
\begin{align*}
z_{im}&=\frac{\log(t_i)-\mu_m}{\sigma_m}, &
Q_{im}&=1-\Phi(z_{im}), &
q_{im}&=\frac{\phi(z_{im})}{t_i\sigma_m},\\
A_i&=\sum_{m=1}^M\pi_{i,m,g_i}Q_{im}, &
B_i&=\sum_{m=1}^M\pi_{i,m,g_i}q_{im}, &
c_i&=c_{g_i}(\bx_i).
\end{align*}
Here $Q_{im}$ and $q_{im}$ are, respectively, the log-normal survival
function and density evaluated at $t_i$. The observed-arm survival function
and density are
\begin{align*}
S_i&=c_i+(1-c_i)A_i, &
f_i&=(1-c_i)B_i.
\end{align*}
Consequently, the observed-data likelihood is
\begin{equation}
L(\bmu,\bsigma,\bbeta,\bGamma,\blambda)
=\prod_{i=1}^N S_i^{\delta_i}f_i^{1-\delta_i},
\qquad
\ell=\sum_{i=1}^N\{\delta_i\log S_i+(1-\delta_i)\log f_i\}.
\label{eq:observed-likelihood}
\end{equation}
Thus, each subject contributes only the likelihood factor corresponding to
the treatment actually received; there is no additional product over arms.

\section{Posterior sampling derivations}\label{supp::scores}

\subsection{Score functions}

For the linear link, let
\begin{align*}
c_g(\bx_i)&=\operatorname{logit}^{-1}(\bx_i^T\blambda_g),\\
\pi_{i,m,g}&=
\frac{\gamma_{m,g}\exp(\bx_i^T\bbeta_m)}
{\sum_{j=1}^M\gamma_{j,g}\exp(\bx_i^T\bbeta_j)}.
\end{align*}
For a component active in arm $g_i$, the likelihood scores are
\begin{align}
\nabla_{\bbeta_m}\ell
&=\sum_{i=1}^N\pi_{i,m,g_i}\bx_i
\left[
\frac{\delta_i(1-c_i)(Q_{im}-A_i)}{S_i}
+\frac{(1-\delta_i)(q_{im}-B_i)}{B_i}
\right],
\label{eq:score-beta}\\
\frac{\partial\ell}{\partial\mu_m}
&=\sum_{i=1}^N\pi_{i,m,g_i}
\left[
\frac{\delta_i(1-c_i)\phi(z_{im})}{\sigma_m S_i}
+\frac{(1-\delta_i)q_{im}\{\log(t_i)-\mu_m\}}
{\sigma_m^2B_i}
\right],
\label{eq:score-mu}\\
\nabla_{\blambda_g}\ell
&=\sum_{i:g_i=g}\bx_i
\left[
\frac{\delta_i c_i(1-c_i)(1-A_i)}{S_i}
-(1-\delta_i)c_i
\right].
\label{eq:score-lambda}
\end{align}
The negative event contribution in \eqref{eq:score-lambda} follows from
$\nabla_{\blambda_g}(1-c_g)=-c_g(1-c_g)\bx_i$.

For the nonlinear link, write
$h_{ik}=\psi(\bx_i^T\btheta_k)$, use $\lambda_{g,k}$ for the cure-output
weight and $\beta_{m,k}$ for the mixture-weight output, and define
\begin{align*}
D^A_{ik}&=\sum_{j=1}^M\pi_{i,j,g_i}\beta_{j,k}(Q_{ij}-A_i),\\
D^B_{ik}&=\sum_{j=1}^M\pi_{i,j,g_i}\beta_{j,k}(q_{ij}-B_i).
\end{align*}
Then
\begin{align}
\nabla_{\btheta_k}\ell
=\sum_{i=1}^N\psi'(\bx_i^T\btheta_k)\bx_i
\bigg[
&\frac{\delta_i}{S_i}
\left\{c_i(1-c_i)\lambda_{g_i,k}(1-A_i)
+(1-c_i)D^A_{ik}\right\} \notag\\
&+(1-\delta_i)
\left\{-c_i\lambda_{g_i,k}+\frac{D^B_{ik}}{B_i}\right\}
\bigg].
\label{eq:score-theta}
\end{align}
Gaussian-prior score terms are added to
Equations~\eqref{eq:score-beta}--\eqref{eq:score-theta}; for example,
$-\bbeta_m/\sigma_\beta^2$, $-\mu_m/\sigma_\mu^2$,
$-\blambda_g/\sigma_\lambda^2$, and
$-\btheta_k/\sigma_\theta^2$.

\section{Bayesian inference}

\subsection{Posterior sampling}\label{posterior}

Continuous parameter blocks are updated by the Metropolis-adjusted
Langevin algorithm (MALA). For a free parameter block $\bw$ and step size
$\epsilon$, the proposal is
\begin{equation*}
\bw'=\bw+\frac{\epsilon}{2}\nabla_{\bw}\log p(\bw\mid\text{rest})
+\sqrt{\epsilon}\,\bz,
\qquad \bz\sim\Normal(\bzero,\bI),
\end{equation*}
and is accepted using the Metropolis--Hastings ratio that includes the
forward and reverse proposal densities. Step sizes are tuned every 200
iterations during burn-in only and are fixed thereafter. The sampler uses
15,000 iterations, discards the first 5,000, and retains every tenth draw,
giving 1,000 posterior draws.

For each $\gamma_{m,g}$, let $\ell_{m,g}^{(1)}$ and
$\ell_{m,g}^{(0)}$ be the observed-data log likelihoods with that indicator
set to 1 and 0, respectively, holding the other quantities fixed. The
conditional inclusion log-odds are
\begin{equation*}
\log\frac{P(\gamma_{m,g}=1\mid\text{rest})}
{P(\gamma_{m,g}=0\mid\text{rest})}
=\log\frac{p_g}{1-p_g}+\ell_{m,g}^{(1)}-\ell_{m,g}^{(0)}.
\end{equation*}
This update is evaluated on the log scale. 
Any update producing $\sum_{m=1}^M\gamma_{m,g}=0$ is assigned probability zero and rejected, ensuring that each treatment arm retains at least one active component.

With $p_g\sim\Beta(c,d)$,
\begin{equation*}
p_g\mid\bGamma\sim
\Beta\left(c+\sum_m\gamma_{m,g},\ d+M-\sum_m\gamma_{m,g}\right).
\end{equation*}

For the inverse-gamma parameterization with density proportional to
$x^{-(a+1)}\exp(-b/x)$, the Gaussian-prior variance updates are
\begin{align*}
\sigma_\mu^2\mid\cdot
&\sim\IG\left(a_\mu+\frac{M}{2},
b_\mu+\frac12\sum_m\mu_m^2\right),\\
\sigma_\beta^2\mid\cdot
&\sim\IG\left(a_\beta+\frac{d_\beta}{2},
b_\beta+\frac12\|\bbeta\|_F^2\right),\\
\sigma_\lambda^2\mid\cdot
&\sim\IG\left(a_\lambda+\frac{d_\lambda}{2},
b_\lambda+\frac12\|\blambda\|_F^2\right),\\
\sigma_\theta^2\mid\cdot
&\sim\IG\left(a_\theta+\frac{d_\theta}{2},
b_\theta+\frac12\|\btheta\|_F^2\right),
\end{align*}
where each $d$ is the number of coefficients in the corresponding parameter
block. 

\section{Survival summaries}

For a finite horizon $h$, the survival probability is
\begin{equation}
P(T>h\mid\bx_i,g)=c_g(\bx_i)
+\{1-c_g(\bx_i)\}\sum_{m=1}^M\pi_{i,m,g}
\left[1-\Phi\left\{\frac{\log h-\mu_m}{\sigma_m}\right\}\right].
\label{eq:survival-probability}
\end{equation}
The corresponding restricted mean survival time is available in closed form:
\begin{align}
\operatorname{RMST}_g(h\mid\bx_i)
&=c_g(\bx_i)h+\{1-c_g(\bx_i)\}
\sum_{m=1}^M\pi_{i,m,g}R_m(h),
\label{eq:rmst}\\
R_m(h)
&=\exp\left(\mu_m+\frac{\sigma_m^2}{2}\right)
\Phi\left\{\frac{\log h-\mu_m-\sigma_m^2}{\sigma_m}\right\}
+h\left[1-\Phi\left\{\frac{\log h-\mu_m}{\sigma_m}\right\}\right].
\end{align}
Treatment contrasts for survival probability or RMST are obtained by
subtracting the corresponding arm-specific quantities at the same
covariate value.

\section{Simulation}
In both simulation cases, a scenario-specific offset is added to the base
cure probability on the probability scale:
\begin{equation*}
c_g(\bx_i)=\min\left\{
\operatorname{logit}^{-1}
\left(\lambda_{g,0}+\bx_i^T\blambda_g\right)
+\operatorname{offset}_g,\ 1\right\}.
\end{equation*}
Thus, $\operatorname{offset}_g$ is not part of the logit-scale linear
predictor. Unless otherwise stated, the elements of $\blambda_g$ are
generated from $\Unif(-2,0)$. We set the $\operatorname{offset}_g$ to 0 and 0.4 to get the desired level of cure fractions. Then, we generate $\gamma_{m,g}\sim\Bernoulli(0.5)$ and $\beta_{m,p}\sim\Normal(0,1)$ for both of the two simulations. These are used to form $\pi_{i,m,g_i}$'s.

\subsection{Well-specified case}

\begin{algorithm}[H]
\caption{Well-specified generation of survival data}
\begin{algorithmic}
\For{\(i \gets 1\) \textbf{to} \(N\)}
    \State \(c_i \gets \min\{\operatorname{logit}^{-1}
    (\lambda_{g_i,0}+\bx_i^T\blambda_{g_i})
    +\operatorname{offset}_{g_i},1\}\)
    \State Draw cure indicator \(U_i\sim\Bernoulli(c_i)\)
    \If{\(U_i=1\)}
        \State \(T_i\gets\infty\)
    \Else
        \State Sample \(m\in\{1,\ldots,M\}\) with probabilities
        \(\pi_{i,m,g_i}\)
        \State Draw \(\log T_i\sim\Normal(\mu_m,\sigma_m^2)\)
    \EndIf
    \State Draw \(C_i\sim\Exp(0.05)\) and set \(h\gets25\)
    \State \(C_i^*\gets\min(C_i,h)\)
    \State \(t_i\gets\min(T_i,C_i^*)\)
    \State \(\delta_i\gets\Ind(T_i>C_i^*)\) \Comment{1 denotes censoring}
\EndFor
\end{algorithmic}
\label{alg:alg_linear}
\end{algorithm}

Table~\ref{tab:rmst_combined} is an additional coefficient-range
sensitivity analysis; the primary simulation uses the range stated above.

\begin{table}[!htpb]
\flushleft
\small
\caption{Additional sensitivity analysis comparing mean squared errors for
RMST between the proposed methods and the Flexible Parametric Cure Models
method, based on 150 replications with
$\lambda_{g,p}\sim\Unif(-1,0)$. The first panel corresponds to
well-specified data generation and the second to misspecified data
generation. Reported errors are medians over replications.}
\centering
\setlength{\tabcolsep}{3pt}
\begin{tabular}{rccc|ccc}
\hline
& \multicolumn{3}{c}{\textbf{High Cure Rate}} 
& \multicolumn{3}{c}{\textbf{Low Cure Rate}} \\
& $\zeta_{21}$ & $\zeta_{31}$ & $\zeta_{32}$ 
& $\zeta_{21}$ & $\zeta_{31}$ & $\zeta_{32}$ \\
\hline
\multicolumn{7}{l}{\textbf{Well-specified Case}} \\
\hline
\textbf{MC Linear} & 5.04 & 0.83 & 4.35 & 21.28 & 1.92 & 21.79 \\
\textbf{MC Nonlinear $K=4$} & 4.86 & 0.76 & 4.36 & 20.91 & 2.05 & 21.12 \\
\textbf{MC Nonlinear $K=6$} & 4.86 & 0.81 & 4.36 & 21.52 & 1.98 & 21.92 \\
\textbf{MC Nonlinear $K=8$} & 4.88 & 0.83 & 4.40 & 21.56 & 1.87 & 21.90 \\
\textbf{MC Nonlinear $K=13$} & 5.04 & 0.93 & 4.34 & 21.47 & 1.82 & 21.92 \\
\textbf{MC Nonlinear $K=18$} & 5.04 & 0.92 & 4.43 & 21.60 & 1.80 & 21.82 \\
\textbf{MC Nonlinear $K=K_{CV}$} & 4.81 & 0.75 & 4.33 & 21.50 & 1.62 & 21.94 \\
\textbf{{\tt flexsurvcure}} & 13.13 & 2.14 & 11.98 & 50.74 & 3.55 & 53.23 \\ 
\hline
\multicolumn{7}{l}{\textbf{Misspecified Case}} \\
\hline
\textbf{MC Linear} & 6.33 & 7.21 & 8.71 & 11.52 & 14.07 & 16.46 \\
\textbf{MC Nonlinear $K=4$} & 6.77 & 7.89 & 8.19 & 12.61 & 12.81 & 16.73 \\
\textbf{MC Nonlinear $K=6$} & 7.30 & 8.31 & 8.96 & 13.63 & 13.16 & 16.71 \\
\textbf{MC Nonlinear $K=8$} & 7.67 & 8.64 & 9.03 & 14.47 & 14.37 & 17.42 \\
\textbf{MC Nonlinear $K=13$} & 8.47 & 9.42 & 9.99 & 15.63 & 15.47 & 18.39 \\
\textbf{MC Nonlinear $K=18$} & 9.99 & 10.68 & 10.03 & 15.38 & 18.72 & 18.99 \\
\textbf{MC Nonlinear $K=K_{CV}$} & 6.75 & 7.34 & 7.91 & 12.32 & 13.40 & 16.64 \\
\textbf{{\tt flexsurvcure}} & 11.40 & 12.14 & 15.28 & 24.50 & 25.99 & 29.06 \\
\hline
\end{tabular}
\label{tab:rmst_combined}
\end{table}

\subsection{Misspecified case}
We first apply k-means clustering with nine centers on the three covariates of the AALL0434 data. This gives us nine clusters. Within each cluster, we fit the log-normal AFT cure models applying the \texttt{flexsurvcure} package. Then, for each of the three simulated treatment arms, outcomes are generated from an arm-specific, covariate-dependent mixture of these nine expert models. Algorithm~\ref{alg:alg_nonlinear} provides additional details.

\begin{figure}[!htpb]
    \centering
    \includegraphics[width = 0.6\linewidth, trim=0cm 0cm 0cm 1cm, clip=true]{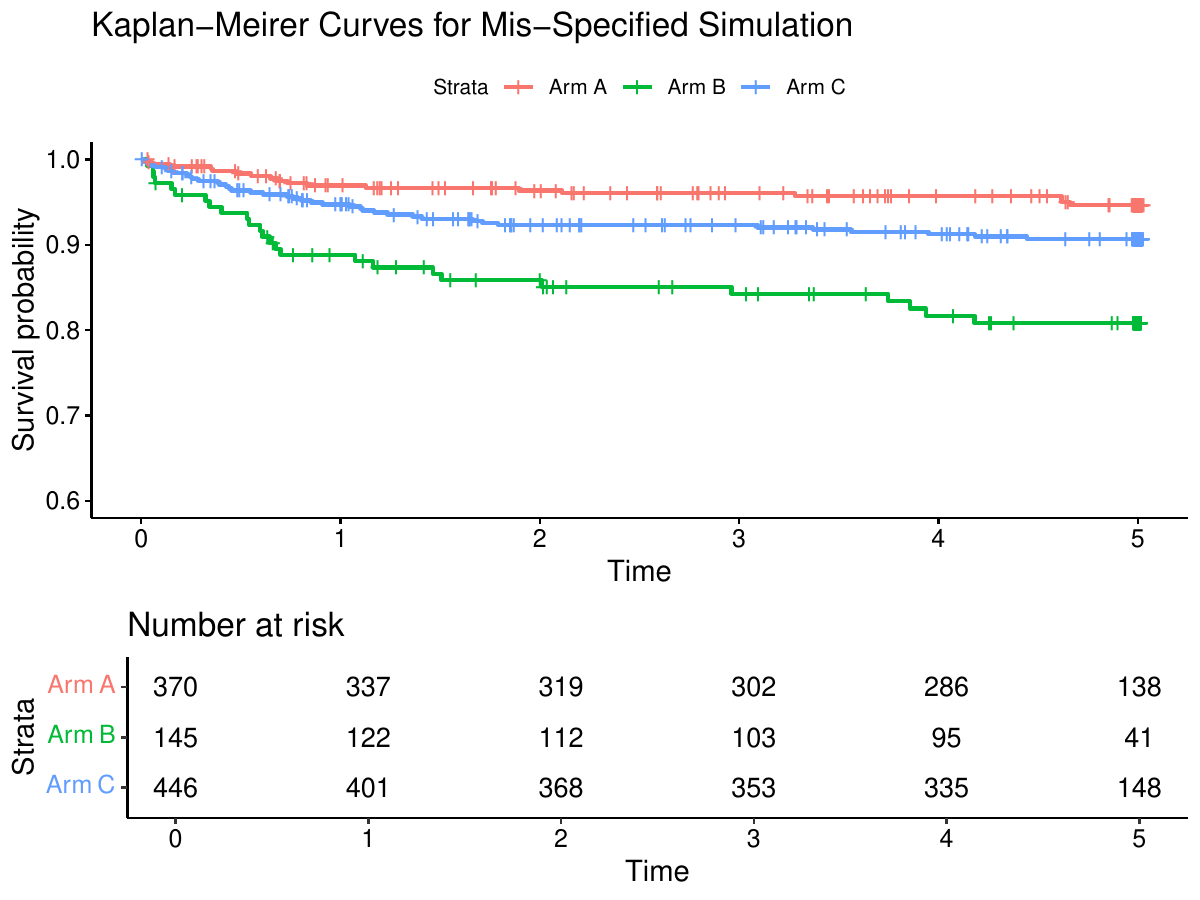}
    \caption{Kaplan--Meier estimates of disease-free survival by treatment
    arm in the misspecified simulation.}
    \label{fig:sim2_curv}
\end{figure}   

\begin{algorithm}[H]
\caption{Misspecified generation of survival data}
\begin{algorithmic}
\For{\(i \gets 1\) \textbf{to} \(N\)}
    \State \(c_i \gets \min\{\operatorname{logit}^{-1}
    (\lambda_{g_i,0}+\bx_i^T\blambda_{g_i})
    +\operatorname{offset}_{g_i},1\}\)
    \State Draw cure indicator \(U_i\sim\Bernoulli(c_i)\)
    \If{\(U_i=1\)}
        \State \(T_i\gets\infty\)
    \Else
        \State Sample \(m\in\{1,\ldots,M\}\) with probabilities
        \(\pi_{i,m,g_i}\)
        \State Draw \(\log T_i\sim\Normal(\mu_{i,m},\sigma_m^2)\), with fitted model parameters from $m$-$th$ model
    \EndIf
    \State Draw \(C_i\sim\Exp(0.05)\) and set \(h\gets5\)
    \State \(C_i^*\gets\min(C_i,h)\)
    \State \(t_i\gets\min(T_i,C_i^*)\)
    \State \(\delta_i\gets\Ind(T_i>C_i^*)\) \Comment{1 denotes censoring}
\EndFor
\end{algorithmic}
\label{alg:alg_nonlinear}
\end{algorithm}

\begin{table}[!htpb]
\flushleft
\caption{Comparison of thresholding-based $s_{p,t}$-values for $\widehat{\bzeta}_{21}$, $\widehat{\bzeta}_{31}$, and $\widehat{\bzeta}_{32}$ for different choices of thresholds ($t$), and predictors in different cure rate settings for the nonlinear simulation ($N=961$, averaged across 150 replications, $K$ selected via cross-validation for each replication). The thresholding levels are given in the first column. The generated outcomes are independent of the CNS Status predictor, shown in italics.}
\centering
\begin{tabular}{r|cccc|cccc}
\hline
& \multicolumn{4}{c|}{\textbf{High Cure Rate}} & \multicolumn{4}{c}{\textbf{Low Cure Rate}} \\
\hline
 \multicolumn{1}{c|}{$\widehat{\bzeta}_{21}$} & Age & WBC & Sex & {\it CNS Status} & Age & WBC & Sex & {\it CNS Status}\\
\hline
  0.00 & 1.00 & 1.00 & 1.00 & 1.00 & 1.00 & 1.00 & 1.00 & 1.00 \\
  0.05 & 0.85 & 0.87 & 0.84 & 0.82 & 0.93 & 0.96 & 0.95 & 0.89 \\
  0.10 & 0.71 & 0.75 & 0.70 & 0.66 & 0.85 & 0.92 & 0.90 & 0.79 \\
  0.15 & 0.60 & 0.64 & 0.58 & 0.52 & 0.78 & 0.88 & 0.84 & 0.70 \\
  0.20 & 0.51 & 0.55 & 0.47 & 0.41 & 0.72 & 0.85 & 0.79 & 0.61 \\
  0.25 & 0.43 & 0.47 & 0.38 & 0.33 & 0.65 & 0.81 & 0.74 & 0.53 \\
  0.30 & 0.36 & 0.41 & 0.31 & 0.26 & 0.60 & 0.77 & 0.69 & 0.45 \\
  0.35 & 0.31 & 0.35 & 0.25 & 0.20 & 0.54 & 0.74 & 0.64 & 0.39 \\
  0.40 & 0.26 & 0.30 & 0.19 & 0.16 & 0.49 & 0.70 & 0.60 & 0.33 \\
\hline
 \multicolumn{1}{c}{$\widehat{\bzeta}_{31}$} \\
\hline
  0.00 & 1.00 & 1.00 & 1.00 & 1.00 & 1.00 & 1.00 & 1.00 & 1.00 \\
  0.05 & 0.84 & 0.87 & 0.94 & 0.77 & 0.94 & 0.99 & 1.00 & 0.92 \\
  0.10 & 0.69 & 0.75 & 0.88 & 0.58 & 0.88 & 0.99 & 1.00 & 0.84 \\
  0.15 & 0.57 & 0.65 & 0.82 & 0.42 & 0.82 & 0.98 & 1.00 & 0.76 \\
  0.20 & 0.47 & 0.57 & 0.76 & 0.30 & 0.77 & 0.98 & 1.00 & 0.68 \\
  0.25 & 0.39 & 0.49 & 0.70 & 0.21 & 0.71 & 0.97 & 0.99 & 0.60 \\
  0.30 & 0.32 & 0.43 & 0.64 & 0.14 & 0.66 & 0.96 & 0.99 & 0.53 \\
  0.35 & 0.26 & 0.37 & 0.57 & 0.10 & 0.60 & 0.96 & 0.99 & 0.47 \\
  0.40 & 0.21 & 0.32 & 0.51 & 0.06 & 0.55 & 0.95 & 0.98 & 0.40 \\
\hline
 \multicolumn{1}{c}{$\widehat{\bzeta}_{32}$} \\
\hline
  0.00 & 1.00 & 1.00 & 1.00 & 1.00 & 1.00 & 1.00 & 1.00 & 1.00 \\
  0.05 & 0.87 & 0.88 & 0.92 & 0.83 & 0.93 & 0.98 & 0.97 & 0.91 \\
  0.10 & 0.74 & 0.78 & 0.84 & 0.67 & 0.86 & 0.95 & 0.94 & 0.82 \\
  0.15 & 0.64 & 0.69 & 0.77 & 0.54 & 0.80 & 0.93 & 0.91 & 0.73 \\
  0.20 & 0.55 & 0.61 & 0.70 & 0.44 & 0.74 & 0.91 & 0.87 & 0.65 \\
  0.25 & 0.47 & 0.54 & 0.63 & 0.35 & 0.68 & 0.89 & 0.84 & 0.58 \\
  0.30 & 0.41 & 0.48 & 0.57 & 0.28 & 0.62 & 0.86 & 0.81 & 0.51 \\
  0.35 & 0.35 & 0.43 & 0.51 & 0.22 & 0.57 & 0.84 & 0.77 & 0.44 \\
  0.40 & 0.30 & 0.38 & 0.45 & 0.17 & 0.52 & 0.82 & 0.74 & 0.38 \\
\hline
\end{tabular}
\label{tab:sim_blp_uni_961}
\end{table}

\begin{table}[!htpb]
\scriptsize
\caption{Comparison of MBLP data partition thresholding-based $s_{p,t}$-values for $\widehat{\bzeta}_{21}$, $\widehat{\bzeta}_{31}$, and $\widehat{\bzeta}_{32}$ for different clusters and predictors in the high cure rate setting for the nonlinear simulation ($N=961$, averaged across 150 replications). The thresholding levels are given in the first column. The generated outcomes are independent of the CNS Status predictor, shown in italics.}
\centering
\scriptsize
\setlength{\tabcolsep}{2pt}
\begin{tabular}{r|cccc|cccc|cccc}
\hline
& \multicolumn{4}{c|}{$\bzeta_{21}$} & \multicolumn{4}{c|}{$\bzeta_{31}$} & \multicolumn{4}{c}{$\bzeta_{32}$} \\
\hline
& Age & WBC & Sex & {\it CNS Status} & Age & WBC & Sex & {\it CNS Status} & Age & WBC & Sex & {\it CNS Status}\\
\hline
& \multicolumn{12}{c}{Cluster 1} \\
\hline
  0.00 & 1.00 & 1.00 & 1.00 & 1.00 & 1.00 & 1.00 & 1.00 & 1.00 & 1.00 & 1.00 & 1.00 & 1.00 \\
  0.05 & 0.87 & 0.87 & 0.85 & 0.81 & 0.86 & 0.87 & 0.93 & 0.78 & 0.88 & 0.89 & 0.91 & 0.83 \\
  0.10 & 0.75 & 0.76 & 0.71 & 0.65 & 0.74 & 0.76 & 0.87 & 0.60 & 0.78 & 0.79 & 0.83 & 0.69 \\
  0.15 & 0.65 & 0.67 & 0.60 & 0.52 & 0.64 & 0.67 & 0.80 & 0.46 & 0.69 & 0.71 & 0.76 & 0.57 \\
  0.20 & 0.56 & 0.59 & 0.49 & 0.42 & 0.55 & 0.59 & 0.74 & 0.36 & 0.61 & 0.64 & 0.69 & 0.47 \\
  0.25 & 0.49 & 0.51 & 0.41 & 0.34 & 0.48 & 0.53 & 0.68 & 0.27 & 0.55 & 0.57 & 0.63 & 0.39 \\
  0.30 & 0.43 & 0.45 & 0.34 & 0.28 & 0.42 & 0.47 & 0.62 & 0.21 & 0.49 & 0.51 & 0.57 & 0.33 \\
  0.35 & 0.37 & 0.40 & 0.28 & 0.22 & 0.36 & 0.41 & 0.56 & 0.16 & 0.44 & 0.46 & 0.51 & 0.27 \\
  0.40 & 0.33 & 0.35 & 0.22 & 0.18 & 0.32 & 0.37 & 0.51 & 0.12 & 0.39 & 0.42 & 0.46 & 0.22 \\
\hline
& \multicolumn{12}{c}{Cluster 2} \\
\hline
  0.00 & 1.00 & 1.00 & 1.00 & 1.00 & 1.00 & 1.00 & 1.00 & 1.00 & 1.00 & 1.00 & 1.00 & 1.00 \\
  0.05 & 0.86 & 0.86 & 0.84 & 0.81 & 0.86 & 0.87 & 0.92 & 0.79 & 0.88 & 0.88 & 0.91 & 0.83 \\
  0.10 & 0.74 & 0.75 & 0.71 & 0.65 & 0.74 & 0.76 & 0.86 & 0.62 & 0.78 & 0.79 & 0.83 & 0.69 \\
  0.15 & 0.64 & 0.65 & 0.59 & 0.52 & 0.64 & 0.67 & 0.80 & 0.49 & 0.69 & 0.70 & 0.75 & 0.58 \\
  0.20 & 0.56 & 0.57 & 0.49 & 0.42 & 0.56 & 0.59 & 0.74 & 0.39 & 0.61 & 0.63 & 0.68 & 0.49 \\
  0.25 & 0.49 & 0.50 & 0.40 & 0.34 & 0.48 & 0.52 & 0.68 & 0.30 & 0.54 & 0.57 & 0.62 & 0.41 \\
  0.30 & 0.43 & 0.44 & 0.33 & 0.28 & 0.42 & 0.46 & 0.62 & 0.24 & 0.48 & 0.51 & 0.56 & 0.34 \\
  0.35 & 0.37 & 0.38 & 0.26 & 0.23 & 0.37 & 0.41 & 0.56 & 0.19 & 0.43 & 0.46 & 0.50 & 0.29 \\
  0.40 & 0.33 & 0.34 & 0.21 & 0.19 & 0.32 & 0.37 & 0.51 & 0.15 & 0.39 & 0.42 & 0.45 & 0.24 \\
\hline
& \multicolumn{12}{c}{Cluster 3} \\
\hline
  0.00 & 1.00 & 1.00 & 1.00 & 1.00 & 1.00 & 1.00 & 1.00 & 1.00 & 1.00 & 1.00 & 1.00 & 1.00 \\
  0.05 & 0.87 & 0.87 & 0.85 & 0.82 & 0.86 & 0.87 & 0.93 & 0.79 & 0.88 & 0.88 & 0.91 & 0.83 \\
  0.10 & 0.75 & 0.76 & 0.72 & 0.67 & 0.74 & 0.76 & 0.87 & 0.61 & 0.78 & 0.79 & 0.84 & 0.69 \\
  0.15 & 0.66 & 0.66 & 0.60 & 0.54 & 0.64 & 0.66 & 0.81 & 0.47 & 0.69 & 0.70 & 0.77 & 0.58 \\
  0.20 & 0.58 & 0.58 & 0.50 & 0.44 & 0.56 & 0.58 & 0.76 & 0.36 & 0.61 & 0.63 & 0.70 & 0.48 \\
  0.25 & 0.51 & 0.51 & 0.42 & 0.36 & 0.49 & 0.51 & 0.70 & 0.28 & 0.55 & 0.56 & 0.64 & 0.40 \\
  0.30 & 0.44 & 0.44 & 0.34 & 0.29 & 0.42 & 0.45 & 0.64 & 0.21 & 0.49 & 0.51 & 0.58 & 0.33 \\
  0.35 & 0.39 & 0.39 & 0.28 & 0.24 & 0.37 & 0.40 & 0.59 & 0.16 & 0.44 & 0.46 & 0.52 & 0.28 \\
  0.40 & 0.35 & 0.34 & 0.23 & 0.20 & 0.32 & 0.36 & 0.53 & 0.13 & 0.39 & 0.41 & 0.47 & 0.23 \\
\hline
& \multicolumn{12}{c}{Cluster 4} \\
\hline
  0.00 & 1.00 & 1.00 & 1.00 & 1.00 & 1.00 & 1.00 & 1.00 & 1.00 & 1.00 & 1.00 & 1.00 & 1.00 \\
  0.05 & 0.87 & 0.87 & 0.85 & 0.81 & 0.86 & 0.87 & 0.92 & 0.78 & 0.88 & 0.89 & 0.90 & 0.83 \\
  0.10 & 0.76 & 0.77 & 0.72 & 0.65 & 0.75 & 0.76 & 0.86 & 0.60 & 0.78 & 0.79 & 0.81 & 0.69 \\
  0.15 & 0.66 & 0.68 & 0.60 & 0.53 & 0.65 & 0.67 & 0.80 & 0.46 & 0.70 & 0.71 & 0.74 & 0.57 \\
  0.20 & 0.58 & 0.60 & 0.50 & 0.43 & 0.57 & 0.59 & 0.74 & 0.35 & 0.62 & 0.64 & 0.67 & 0.47 \\
  0.25 & 0.51 & 0.53 & 0.42 & 0.35 & 0.50 & 0.53 & 0.69 & 0.27 & 0.55 & 0.58 & 0.60 & 0.40 \\
  0.30 & 0.45 & 0.47 & 0.34 & 0.28 & 0.44 & 0.47 & 0.63 & 0.21 & 0.49 & 0.53 & 0.54 & 0.33 \\
  0.35 & 0.40 & 0.42 & 0.28 & 0.23 & 0.39 & 0.42 & 0.57 & 0.16 & 0.44 & 0.48 & 0.49 & 0.28 \\
  0.40 & 0.35 & 0.38 & 0.23 & 0.19 & 0.34 & 0.38 & 0.52 & 0.13 & 0.40 & 0.43 & 0.44 & 0.24 \\
\hline
\end{tabular}
\label{tab:sim_blp_part_hr_961}
\end{table}

\begin{table}[!htpb]
\scriptsize
\caption{Comparison of MBLP data partition thresholding-based $s_{p,t}$-values for $\widehat{\bzeta}_{21}$, $\widehat{\bzeta}_{31}$, and $\widehat{\bzeta}_{32}$ for different clusters and predictors in the low cure rate setting for the nonlinear simulation ($N=961$, averaged across 150 replications). The thresholding levels are given in the first column. The generated outcomes are independent of the CNS Status predictor, shown in italics.}
\centering
\scriptsize
\setlength{\tabcolsep}{2pt}
\begin{tabular}{r|cccc|cccc|cccc}
\hline
& \multicolumn{4}{c|}{$\bzeta_{21}$} & \multicolumn{4}{c|}{$\bzeta_{31}$} & \multicolumn{4}{c}{$\bzeta_{32}$} \\
\hline
& Age & WBC & Sex & {\it CNS Status} & Age & WBC & Sex & {\it CNS Status} & Age & WBC & Sex & {\it CNS Status}\\
\hline
& \multicolumn{12}{c}{Cluster 1} \\
\hline
  0.00 & 1.00 & 1.00 & 1.00 & 1.00 & 1.00 & 1.00 & 1.00 & 1.00 & 1.00 & 1.00 & 1.00 & 1.00 \\
  0.05 & 0.94 & 0.96 & 0.95 & 0.90 & 0.95 & 0.98 & 0.99 & 0.91 & 0.94 & 0.97 & 0.96 & 0.90 \\
  0.10 & 0.88 & 0.92 & 0.90 & 0.80 & 0.91 & 0.96 & 0.99 & 0.82 & 0.88 & 0.93 & 0.92 & 0.81 \\
  0.15 & 0.83 & 0.88 & 0.84 & 0.71 & 0.87 & 0.95 & 0.98 & 0.73 & 0.83 & 0.90 & 0.89 & 0.72 \\
  0.20 & 0.78 & 0.84 & 0.79 & 0.63 & 0.82 & 0.93 & 0.97 & 0.66 & 0.78 & 0.87 & 0.85 & 0.65 \\
  0.25 & 0.73 & 0.81 & 0.74 & 0.56 & 0.78 & 0.91 & 0.97 & 0.58 & 0.73 & 0.84 & 0.81 & 0.57 \\
  0.30 & 0.68 & 0.77 & 0.70 & 0.49 & 0.74 & 0.90 & 0.96 & 0.51 & 0.68 & 0.81 & 0.77 & 0.51 \\
  0.35 & 0.64 & 0.74 & 0.65 & 0.43 & 0.70 & 0.88 & 0.95 & 0.45 & 0.64 & 0.79 & 0.73 & 0.45 \\
  0.40 & 0.59 & 0.70 & 0.60 & 0.37 & 0.67 & 0.86 & 0.93 & 0.40 & 0.60 & 0.76 & 0.69 & 0.40 \\
\hline
& \multicolumn{12}{c}{Cluster 2} \\
\hline
  0.00 & 1.00 & 1.00 & 1.00 & 1.00 & 1.00 & 1.00 & 1.00 & 1.00 & 1.00 & 1.00 & 1.00 & 1.00 \\
  0.05 & 0.93 & 0.96 & 0.94 & 0.89 & 0.95 & 0.98 & 0.99 & 0.90 & 0.94 & 0.97 & 0.97 & 0.88 \\
  0.10 & 0.87 & 0.91 & 0.88 & 0.79 & 0.91 & 0.96 & 0.99 & 0.81 & 0.88 & 0.93 & 0.93 & 0.79 \\
  0.15 & 0.81 & 0.88 & 0.82 & 0.70 & 0.86 & 0.95 & 0.98 & 0.73 & 0.83 & 0.91 & 0.90 & 0.70 \\
  0.20 & 0.76 & 0.84 & 0.76 & 0.62 & 0.82 & 0.93 & 0.97 & 0.65 & 0.78 & 0.88 & 0.86 & 0.62 \\
  0.25 & 0.70 & 0.80 & 0.71 & 0.55 & 0.78 & 0.91 & 0.96 & 0.58 & 0.73 & 0.85 & 0.83 & 0.55 \\
  0.30 & 0.66 & 0.76 & 0.65 & 0.48 & 0.74 & 0.90 & 0.95 & 0.51 & 0.68 & 0.82 & 0.79 & 0.49 \\
  0.35 & 0.61 & 0.73 & 0.60 & 0.42 & 0.70 & 0.88 & 0.94 & 0.45 & 0.64 & 0.80 & 0.76 & 0.43 \\
  0.40 & 0.57 & 0.70 & 0.56 & 0.37 & 0.67 & 0.86 & 0.93 & 0.39 & 0.60 & 0.77 & 0.72 & 0.38 \\
\hline
& \multicolumn{12}{c}{Cluster 3} \\
\hline
  0.00 & 1.00 & 1.00 & 1.00 & 1.00 & 1.00 & 1.00 & 1.00 & 1.00 & 1.00 & 1.00 & 1.00 & 1.00 \\
  0.05 & 0.94 & 0.96 & 0.95 & 0.90 & 0.96 & 0.98 & 1.00 & 0.91 & 0.95 & 0.97 & 0.96 & 0.90 \\
  0.10 & 0.89 & 0.92 & 0.90 & 0.81 & 0.92 & 0.97 & 0.99 & 0.83 & 0.89 & 0.94 & 0.93 & 0.81 \\
  0.15 & 0.83 & 0.89 & 0.85 & 0.72 & 0.87 & 0.95 & 0.99 & 0.75 & 0.84 & 0.91 & 0.89 & 0.73 \\
  0.20 & 0.78 & 0.85 & 0.80 & 0.63 & 0.83 & 0.94 & 0.98 & 0.68 & 0.80 & 0.88 & 0.85 & 0.65 \\
  0.25 & 0.73 & 0.81 & 0.75 & 0.56 & 0.79 & 0.92 & 0.97 & 0.60 & 0.75 & 0.86 & 0.82 & 0.59 \\
  0.30 & 0.68 & 0.78 & 0.70 & 0.49 & 0.76 & 0.91 & 0.96 & 0.53 & 0.71 & 0.83 & 0.78 & 0.53 \\
  0.35 & 0.63 & 0.74 & 0.65 & 0.43 & 0.72 & 0.89 & 0.95 & 0.47 & 0.66 & 0.80 & 0.74 & 0.47 \\
  0.40 & 0.59 & 0.71 & 0.61 & 0.38 & 0.68 & 0.88 & 0.94 & 0.41 & 0.62 & 0.78 & 0.70 & 0.42 \\
\hline
& \multicolumn{12}{c}{Cluster 4} \\
\hline
  0.00 & 1.00 & 1.00 & 1.00 & 1.00 & 1.00 & 1.00 & 1.00 & 1.00 & 1.00 & 1.00 & 1.00 & 1.00 \\
  0.05 & 0.94 & 0.96 & 0.94 & 0.91 & 0.95 & 0.98 & 0.99 & 0.92 & 0.94 & 0.97 & 0.96 & 0.92 \\
  0.10 & 0.89 & 0.92 & 0.89 & 0.81 & 0.91 & 0.96 & 0.98 & 0.85 & 0.89 & 0.94 & 0.92 & 0.84 \\
  0.15 & 0.83 & 0.88 & 0.84 & 0.73 & 0.86 & 0.95 & 0.97 & 0.77 & 0.83 & 0.90 & 0.88 & 0.77 \\
  0.20 & 0.78 & 0.84 & 0.78 & 0.65 & 0.82 & 0.93 & 0.96 & 0.70 & 0.78 & 0.88 & 0.84 & 0.71 \\
  0.25 & 0.73 & 0.81 & 0.73 & 0.58 & 0.78 & 0.91 & 0.95 & 0.64 & 0.74 & 0.85 & 0.80 & 0.64 \\
  0.30 & 0.68 & 0.77 & 0.68 & 0.52 & 0.74 & 0.89 & 0.93 & 0.57 & 0.69 & 0.82 & 0.76 & 0.58 \\
  0.35 & 0.64 & 0.74 & 0.64 & 0.45 & 0.70 & 0.88 & 0.92 & 0.51 & 0.65 & 0.79 & 0.72 & 0.53 \\
  0.40 & 0.60 & 0.71 & 0.59 & 0.40 & 0.67 & 0.86 & 0.90 & 0.45 & 0.61 & 0.76 & 0.69 & 0.48 \\
\hline
\end{tabular}
\label{tab:sim_blp_part_lr_961}
\end{table}

\newpage
\section{AALL0434 data analysis}

\begin{table}
\centering
\small
\caption{Baseline characteristics of the AALL0434 analysis sample by
treatment arm. Continuous variables are reported as mean (SD), and
categorical variables as $n$ (\%). The descriptive summaries are reported on the original covariate scales; the rescaled and dummy-encoded covariates are used for model fitting.}
\vspace{-0.25cm}
\resizebox{\linewidth}{!}{%
\begin{tabular}{l|ccccc}
\textbf{Variable} & \textbf{All} & \textbf{C-MTX} & \textbf{C-MTX/nelarabine} & \textbf{HDMTX} & \textbf{HDMTX/nelarabine} \\
\hline
\textbf{N} & 1022 & 370 (36.2\%) & 145 (14.2\%) & 362 (35.4\%) & 145 (14.2\%) \\
\textbf{Age (years)} & 9.9 (5.3) & 9.7 (5.3) & 10.5 (5.8) & 9.5 (5.1) & 10.9 (5.5) \\
\textbf{WBC count ($10^3$ cells/$\mu$L)} & 147.9	(180.3) & 146.1	(177.2) & 153.1	(182.8) & 144.2	(168.1) & 156.7	(213.9) \\
\hline
\textbf{Sex} & & & & & \\
\hspace{3mm} Female & 248 (24.3\%) & 86 (23.2\%) & 37 (25.5\%) & 88 (24.3\%) & 37 (25.5\%) \\
\hspace{3mm} Male &  774 (75.7\%) &  284 (76.8\%) & 108 (74.5\%) & 274 (75.7\%) & 108 (74.5\%) \\
\hline
\textbf{Risk Group} & & & & & \\
\hspace{3mm} High risk T-ALL        & 210 (20.5\%) & 55 (14.9\%) & 51 (35.2\%) & 52 (14.4\%) & 52 (35.9\%) \\
\hspace{3mm} Intermediate risk T-ALL & 703 (68.8\%) & 261 (70.5\%) & 94 (64.8\%) & 255 (70.4\%) & 93 (64.1\%) \\
\hspace{3mm} Low risk T-ALL          & 109 (10.7\%) & 54 (14.6\%) & 0 (0\%) & 55 (15.2\%) & 0 (0\%) \\
\hline
\textbf{CNS Status} & & & & & \\
\hspace{3mm} 1 & 810 (79.3\%) & 286 (77.3\%) & 113 (77.9\%) & 295 (81.5\%) & 116 (80\%) \\
\hspace{3mm} 2 & 212 (20.7\%) & 84 (22.7\%) & 32 (22.1\%) & 67 (18.5\%) & 29 (20\%) \\
\end{tabular}%
}
\label{tab:aall_dem}
\end{table}
\begin{figure}[!htpb]
    \centering
    \includegraphics[width = 0.6\linewidth]{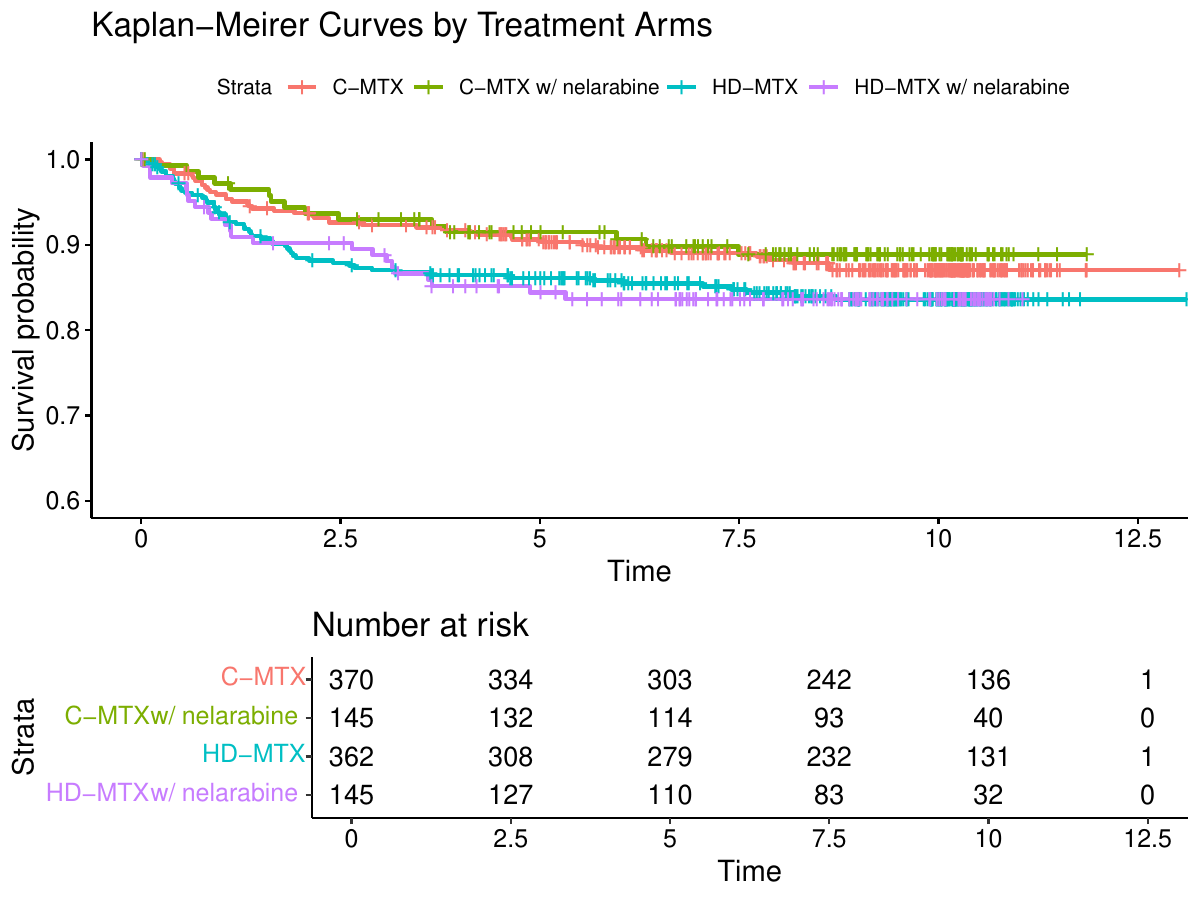}
    \caption{Kaplan--Meier estimates of disease-free survival by treatment
    arm in the AALL0434 analysis sample.}
    \label{fig:aall0434_curv}
\end{figure}

\begin{table}[!htpb]
\centering
\scriptsize
\setlength{\tabcolsep}{2pt}
\caption{MBLP thresholding-based $s_{p,t}$-values for
$\widehat{\bzeta}_{41}$, $\widehat{\bzeta}_{32}$, and
$\widehat{\bzeta}_{42}$ under different thresholds $t$ in the AALL0434
analysis. Results are shown for the full data and two data partitions. The
low-risk columns are omitted because low-risk patients were not eligible for
nelarabine.}
\label{tab:aall_mblp}
\begin{tabular}{c*{15}{c}}
\hline
& \multicolumn{5}{c}{\textbf{Full data}}
& \multicolumn{5}{c}{\textbf{Partition 1}}
& \multicolumn{5}{c}{\textbf{Partition 2}} \\
\cline{2-6}\cline{7-11}\cline{12-16}
\textbf{Threshold}
& Age & \makecell{CNS\\status} & Sex & \makecell{WBC\\count} & \makecell{High\\risk}
& Age & \makecell{CNS\\status} & Sex & \makecell{WBC\\count} & \makecell{High\\risk}
& Age & \makecell{CNS\\status} & Sex & \makecell{WBC\\count} & \makecell{High\\risk} \\
\hline
& \multicolumn{15}{c}{$\widehat{\bzeta}_{41}$} \\
\hline
0.05 & 0.97 & 0.94 & 0.95 & 0.97 & 0.94 & 0.94 & 0.94 & 0.95 & 0.97 & 0.94 & 0.96 & 0.94 & 0.95 & 0.97 & 0.93 \\
0.15 & 0.89 & 0.85 & 0.87 & 0.91 & 0.80 & 0.86 & 0.84 & 0.84 & 0.90 & 0.83 & 0.87 & 0.86 & 0.87 & 0.90 & 0.81 \\
0.25 & 0.81 & 0.75 & 0.77 & 0.86 & 0.67 & 0.78 & 0.75 & 0.74 & 0.83 & 0.69 & 0.79 & 0.76 & 0.79 & 0.82 & 0.68 \\
0.35 & 0.76 & 0.64 & 0.68 & 0.78 & 0.54 & 0.71 & 0.65 & 0.65 & 0.76 & 0.58 & 0.71 & 0.67 & 0.70 & 0.76 & 0.55 \\
0.40 & 0.72 & 0.59 & 0.62 & 0.75 & 0.50 & 0.68 & 0.59 & 0.59 & 0.73 & 0.53 & 0.68 & 0.61 & 0.65 & 0.72 & 0.50 \\
0.50 & 0.65 & 0.51 & 0.53 & 0.70 & 0.40 & 0.60 & 0.50 & 0.51 & 0.67 & 0.42 & 0.60 & 0.52 & 0.56 & 0.64 & 0.40 \\
0.60 & 0.59 & 0.43 & 0.43 & 0.63 & 0.31 & 0.54 & 0.42 & 0.42 & 0.60 & 0.34 & 0.55 & 0.45 & 0.47 & 0.57 & 0.32 \\
0.70 & 0.52 & 0.35 & 0.34 & 0.58 & 0.24 & 0.48 & 0.36 & 0.35 & 0.54 & 0.26 & 0.48 & 0.38 & 0.38 & 0.51 & 0.25 \\
0.80 & 0.46 & 0.29 & 0.28 & 0.52 & 0.18 & 0.43 & 0.29 & 0.28 & 0.49 & 0.21 & 0.43 & 0.31 & 0.31 & 0.46 & 0.19 \\
0.90 & 0.41 & 0.23 & 0.21 & 0.47 & 0.15 & 0.38 & 0.24 & 0.21 & 0.44 & 0.17 & 0.37 & 0.26 & 0.24 & 0.41 & 0.14 \\
1.00 & 0.36 & 0.18 & 0.16 & 0.44 & 0.10 & 0.33 & 0.20 & 0.15 & 0.39 & 0.14 & 0.33 & 0.21 & 0.18 & 0.38 & 0.10 \\
\hline
& \multicolumn{15}{c}{$\widehat{\bzeta}_{32}$} \\
\hline
0.05 & 0.94 & 0.94 & 0.92 & 0.99 & 0.97 & 0.96 & 0.96 & 0.92 & 0.99 & 0.97 & 0.94 & 0.92 & 0.91 & 0.99 & 0.97 \\
0.15 & 0.83 & 0.82 & 0.74 & 0.97 & 0.91 & 0.85 & 0.88 & 0.80 & 0.98 & 0.90 & 0.80 & 0.80 & 0.73 & 0.98 & 0.91 \\
0.25 & 0.72 & 0.72 & 0.58 & 0.96 & 0.85 & 0.76 & 0.80 & 0.64 & 0.96 & 0.83 & 0.67 & 0.67 & 0.58 & 0.96 & 0.86 \\
0.35 & 0.62 & 0.63 & 0.45 & 0.95 & 0.79 & 0.68 & 0.74 & 0.50 & 0.96 & 0.77 & 0.57 & 0.56 & 0.44 & 0.95 & 0.80 \\
0.40 & 0.58 & 0.58 & 0.38 & 0.94 & 0.77 & 0.64 & 0.69 & 0.46 & 0.95 & 0.75 & 0.52 & 0.50 & 0.39 & 0.95 & 0.77 \\
0.50 & 0.49 & 0.49 & 0.29 & 0.93 & 0.70 & 0.55 & 0.60 & 0.37 & 0.94 & 0.69 & 0.43 & 0.42 & 0.28 & 0.93 & 0.70 \\
0.60 & 0.41 & 0.41 & 0.21 & 0.91 & 0.62 & 0.48 & 0.52 & 0.28 & 0.92 & 0.62 & 0.35 & 0.30 & 0.21 & 0.92 & 0.64 \\
0.70 & 0.35 & 0.32 & 0.15 & 0.89 & 0.55 & 0.42 & 0.45 & 0.21 & 0.91 & 0.56 & 0.29 & 0.22 & 0.15 & 0.90 & 0.57 \\
0.80 & 0.28 & 0.23 & 0.10 & 0.87 & 0.49 & 0.36 & 0.37 & 0.14 & 0.89 & 0.49 & 0.23 & 0.15 & 0.10 & 0.88 & 0.49 \\
0.90 & 0.23 & 0.17 & 0.07 & 0.85 & 0.42 & 0.30 & 0.30 & 0.10 & 0.86 & 0.45 & 0.20 & 0.12 & 0.07 & 0.86 & 0.41 \\
1.00 & 0.18 & 0.13 & 0.04 & 0.82 & 0.33 & 0.27 & 0.24 & 0.08 & 0.84 & 0.38 & 0.15 & 0.09 & 0.05 & 0.83 & 0.34 \\
\hline
& \multicolumn{15}{c}{$\widehat{\bzeta}_{42}$} \\
\hline
0.05 & 0.95 & 0.93 & 0.92 & 0.97 & 0.94 & 0.94 & 0.94 & 0.92 & 0.96 & 0.95 & 0.94 & 0.93 & 0.92 & 0.96 & 0.95 \\
0.15 & 0.84 & 0.78 & 0.76 & 0.90 & 0.83 & 0.84 & 0.80 & 0.79 & 0.88 & 0.84 & 0.83 & 0.78 & 0.76 & 0.90 & 0.84 \\
0.25 & 0.76 & 0.64 & 0.62 & 0.81 & 0.73 & 0.74 & 0.67 & 0.65 & 0.80 & 0.76 & 0.74 & 0.65 & 0.62 & 0.82 & 0.74 \\
0.35 & 0.66 & 0.53 & 0.49 & 0.73 & 0.64 & 0.66 & 0.55 & 0.53 & 0.72 & 0.66 & 0.64 & 0.54 & 0.49 & 0.76 & 0.65 \\
0.40 & 0.61 & 0.48 & 0.43 & 0.69 & 0.60 & 0.61 & 0.49 & 0.46 & 0.69 & 0.61 & 0.60 & 0.48 & 0.42 & 0.72 & 0.61 \\
0.50 & 0.53 & 0.38 & 0.32 & 0.62 & 0.52 & 0.53 & 0.40 & 0.37 & 0.63 & 0.51 & 0.52 & 0.38 & 0.31 & 0.66 & 0.53 \\
0.60 & 0.47 & 0.29 & 0.22 & 0.56 & 0.44 & 0.46 & 0.32 & 0.28 & 0.56 & 0.42 & 0.45 & 0.30 & 0.22 & 0.62 & 0.44 \\
0.70 & 0.40 & 0.22 & 0.15 & 0.50 & 0.35 & 0.41 & 0.25 & 0.20 & 0.51 & 0.35 & 0.39 & 0.23 & 0.15 & 0.55 & 0.36 \\
0.80 & 0.34 & 0.17 & 0.11 & 0.44 & 0.28 & 0.36 & 0.19 & 0.14 & 0.46 & 0.28 & 0.32 & 0.18 & 0.11 & 0.50 & 0.29 \\
0.90 & 0.30 & 0.13 & 0.07 & 0.39 & 0.22 & 0.31 & 0.15 & 0.10 & 0.41 & 0.22 & 0.28 & 0.15 & 0.07 & 0.45 & 0.23 \\
1.00 & 0.26 & 0.11 & 0.05 & 0.34 & 0.18 & 0.27 & 0.12 & 0.07 & 0.36 & 0.18 & 0.23 & 0.11 & 0.05 & 0.40 & 0.18 \\
\hline
\end{tabular}
\end{table}

\begin{figure}[!htpb]
  \centering
  \begin{subfigure}[b]{0.45\textwidth}
    \includegraphics[width=\textwidth]{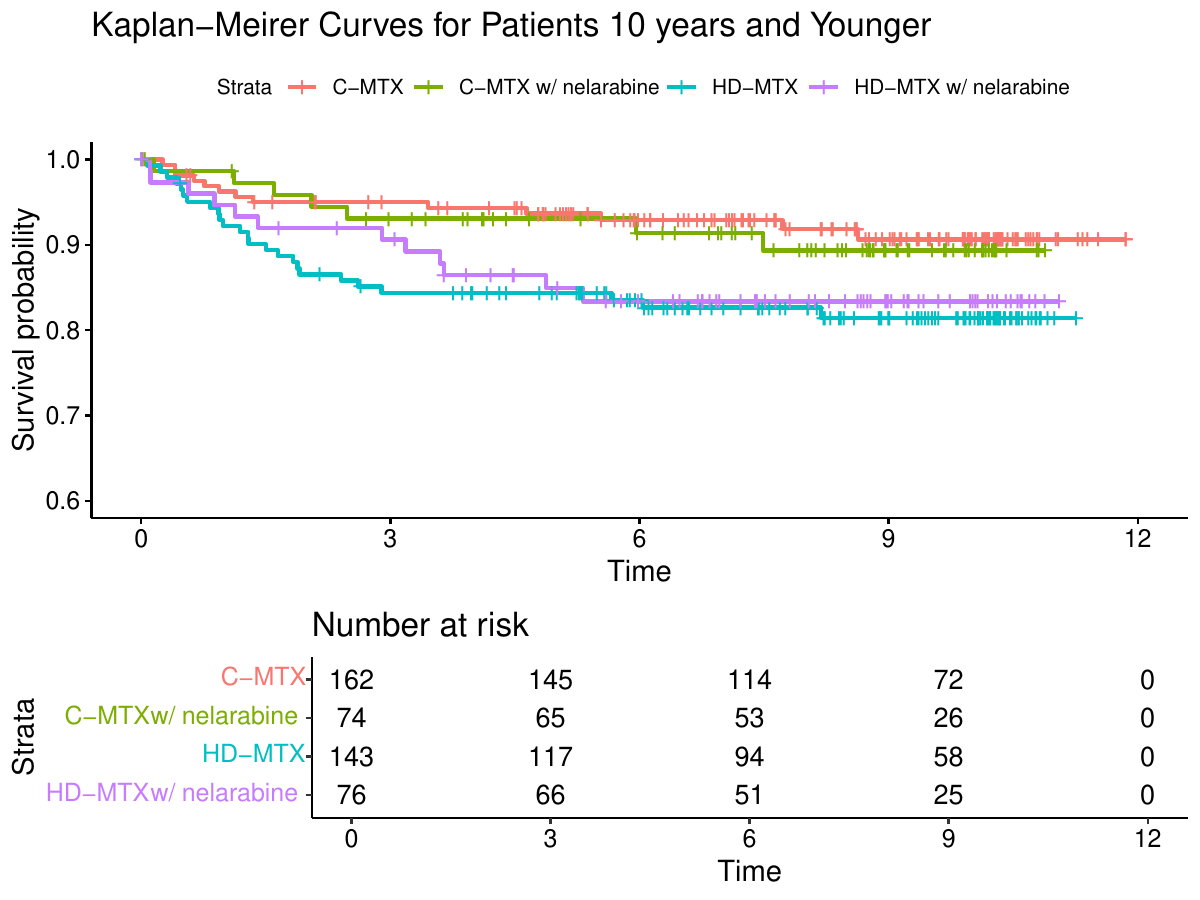}
  \end{subfigure}
  \quad
  \begin{subfigure}[b]{0.45\textwidth}
    \includegraphics[width=\textwidth]{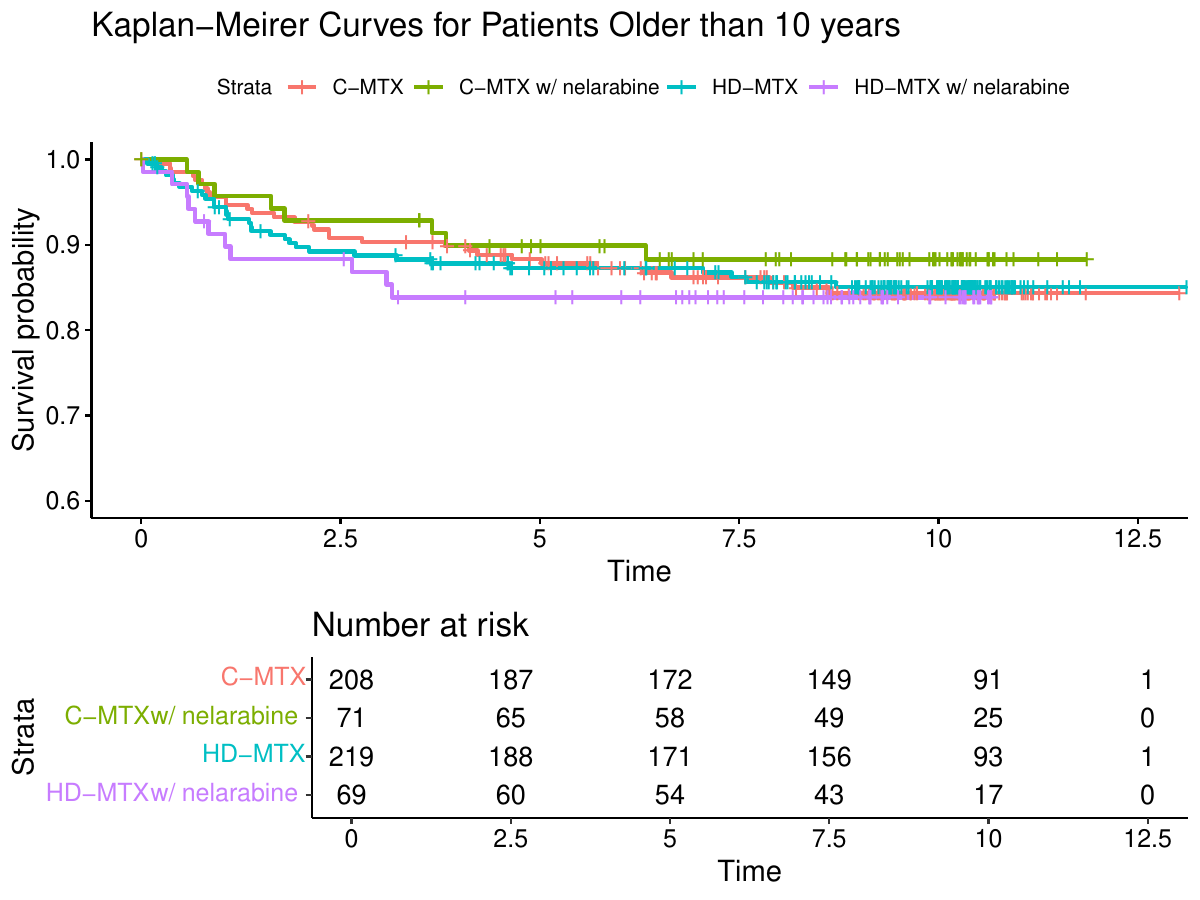}
  \end{subfigure}
  \caption{Kaplan--Meier estimates stratified by age.}
  \label{fig:km_age}
\end{figure}

\begin{figure}[!htpb]
  \centering
  \begin{subfigure}[b]{0.45\textwidth}
    \includegraphics[width=\textwidth]{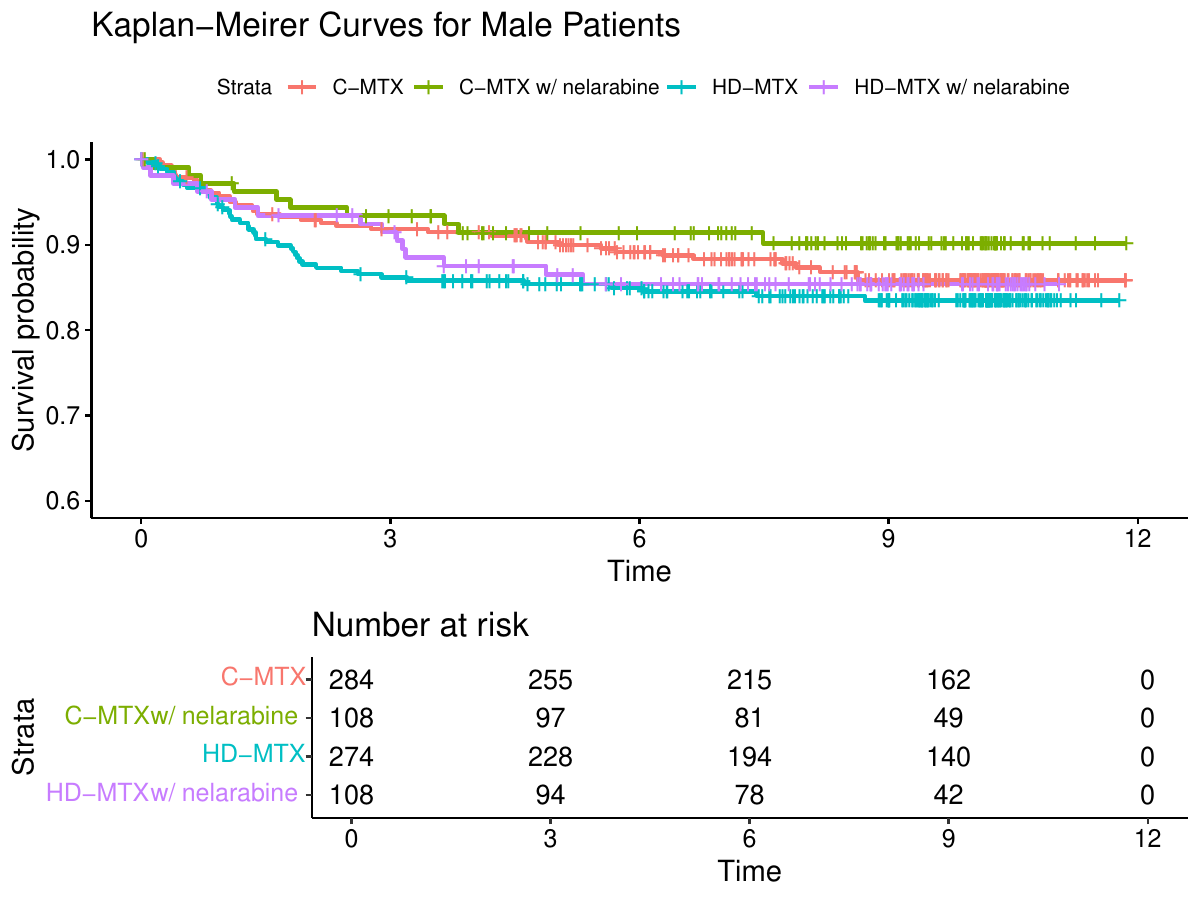}
  \end{subfigure}
  \quad
  \begin{subfigure}[b]{0.45\textwidth}
    \includegraphics[width=\textwidth]{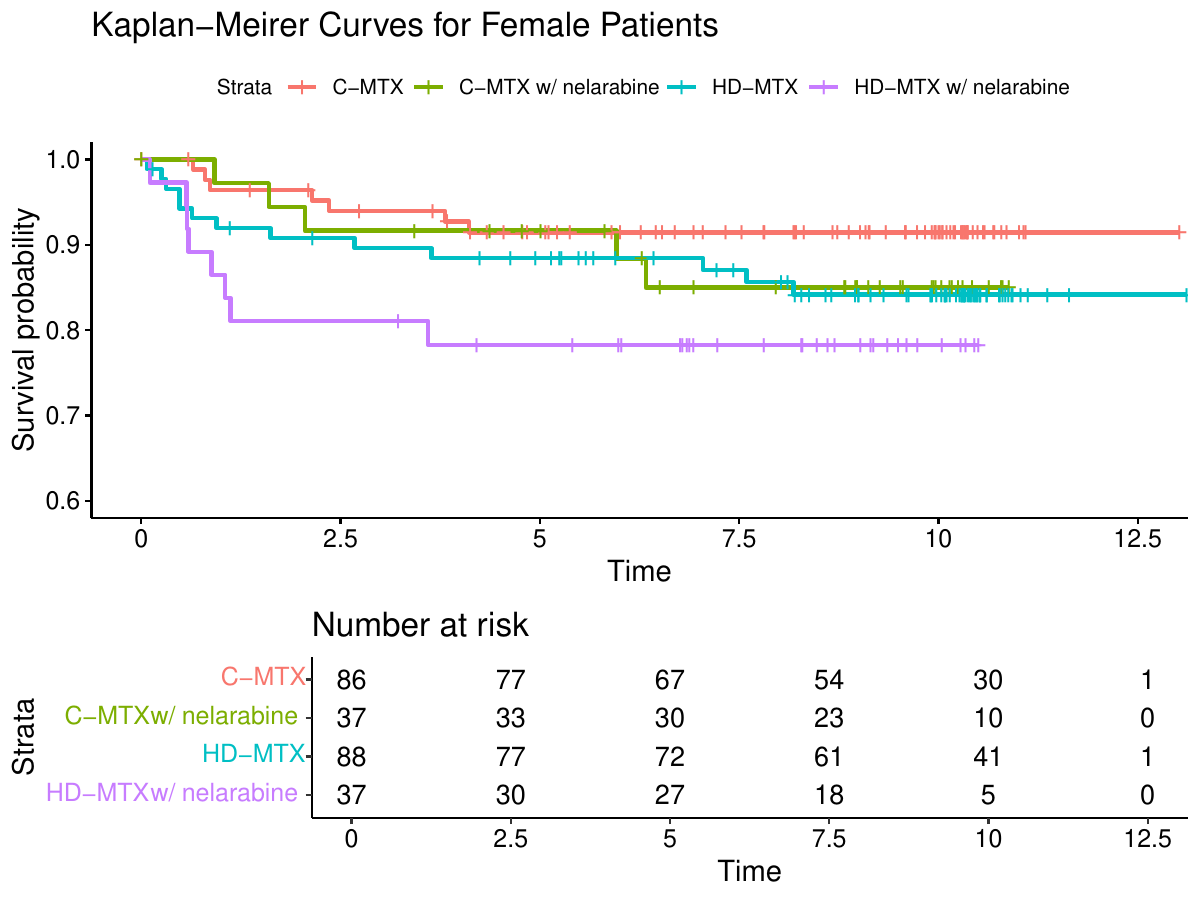}
  \end{subfigure}
  \caption{Kaplan--Meier estimates stratified by sex.}
  \label{fig:km_sex}
\end{figure}

\begin{figure}[!htpb]
  \centering
  \begin{subfigure}[b]{0.45\textwidth}
    \includegraphics[width=\textwidth]{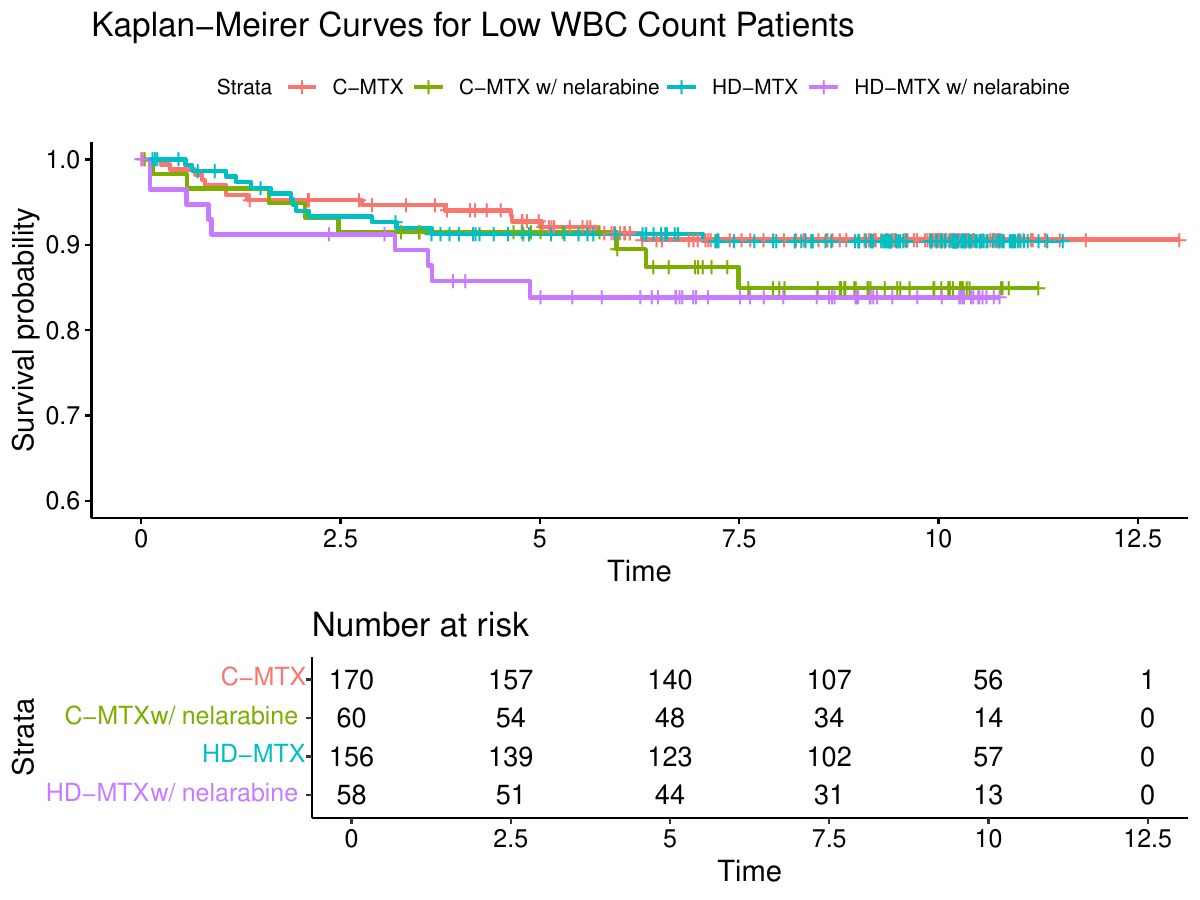}
  \end{subfigure}
  \quad
  \begin{subfigure}[b]{0.45\textwidth}
    \includegraphics[width=\textwidth]{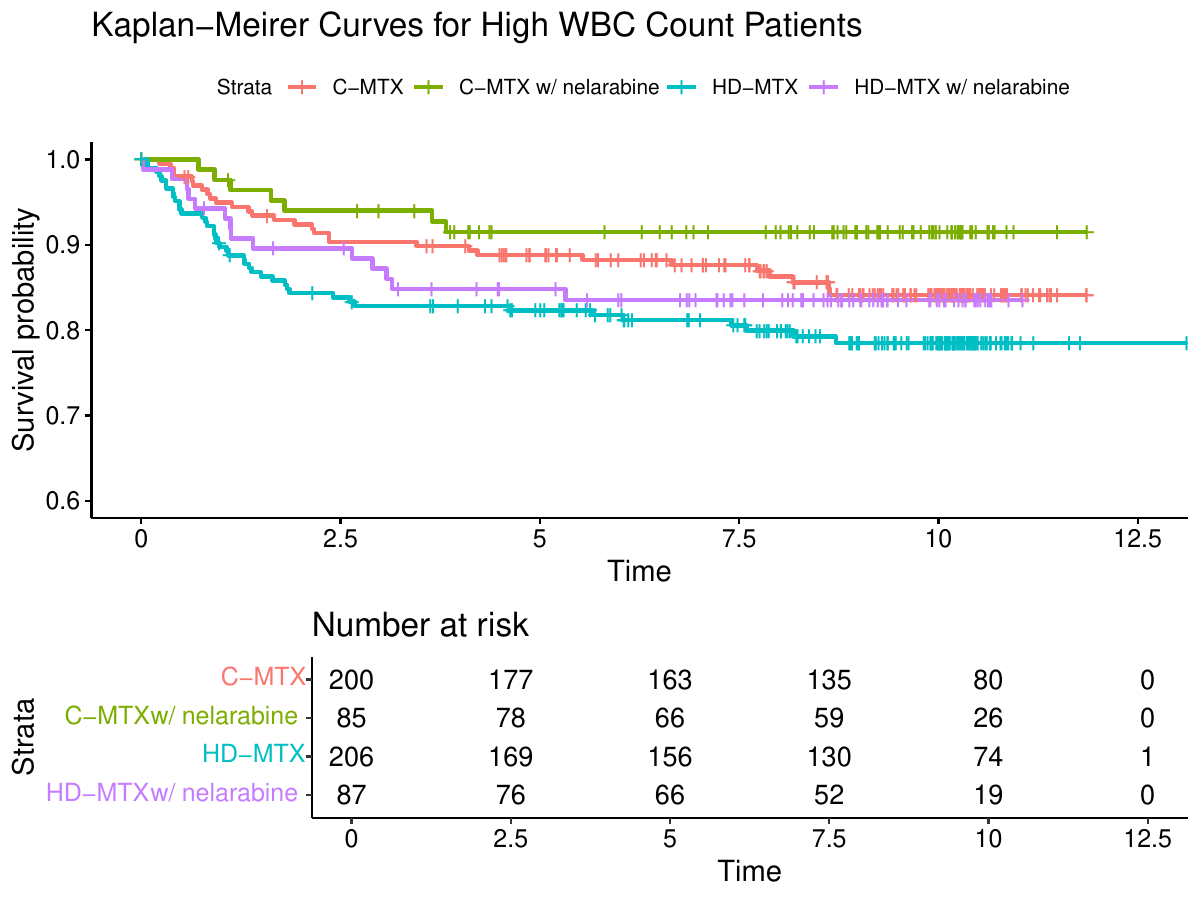}
  \end{subfigure}
  \caption{Kaplan--Meier estimates stratified by WBC count; high WBC is
  defined as at least 50,000 cells/$\mu$L.}
  \label{fig:km_wbc}
\end{figure}

\begin{figure}[!htpb]
  \centering
  \begin{subfigure}[b]{0.45\textwidth}
    \includegraphics[width=\textwidth]{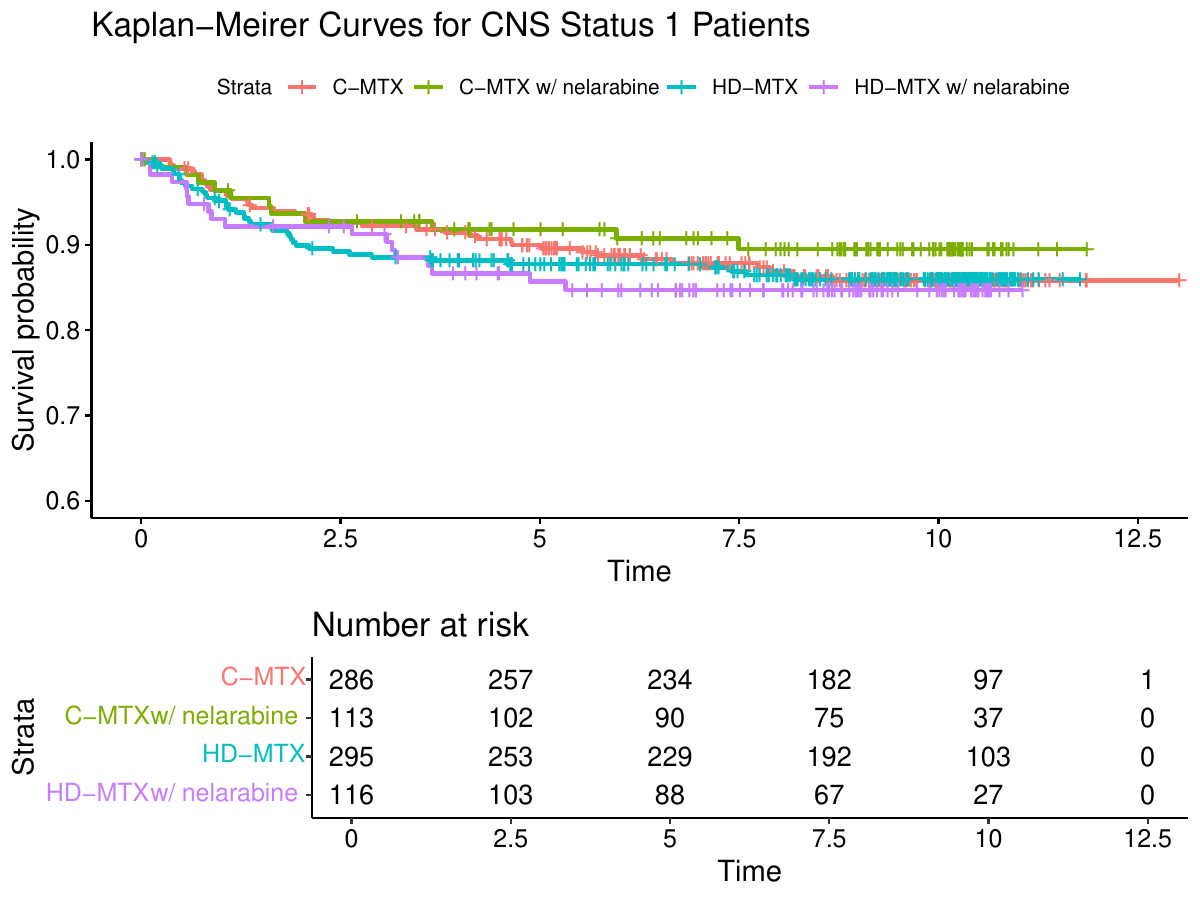}
  \end{subfigure}
  \quad
  \begin{subfigure}[b]{0.45\textwidth}
    \includegraphics[width=\textwidth]{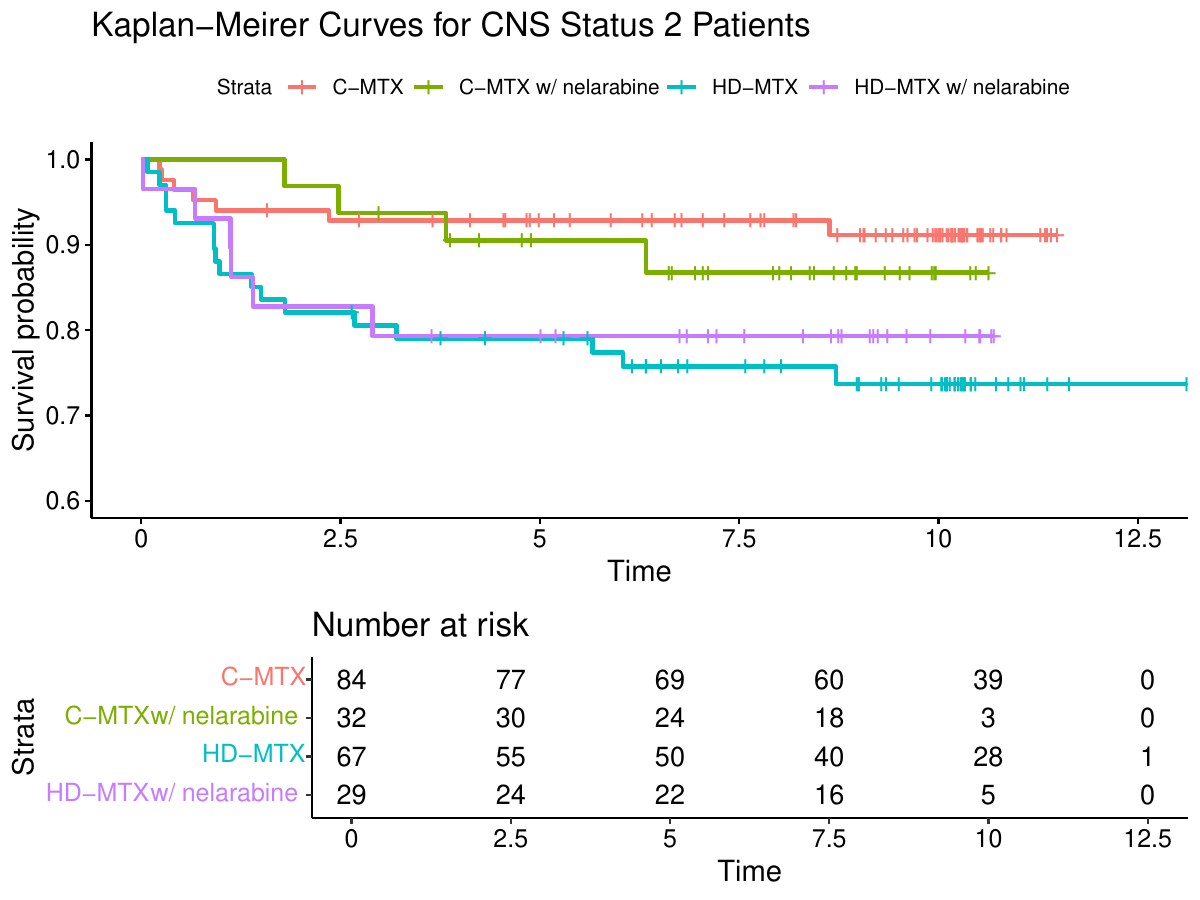}
  \end{subfigure}
  \caption{Kaplan--Meier estimates stratified by CNS status.}
  \label{fig:km_cns}
\end{figure}

\begin{figure}[!htpb]
  \centering
  \begin{subfigure}[b]{0.45\textwidth}
    \includegraphics[width=\textwidth]{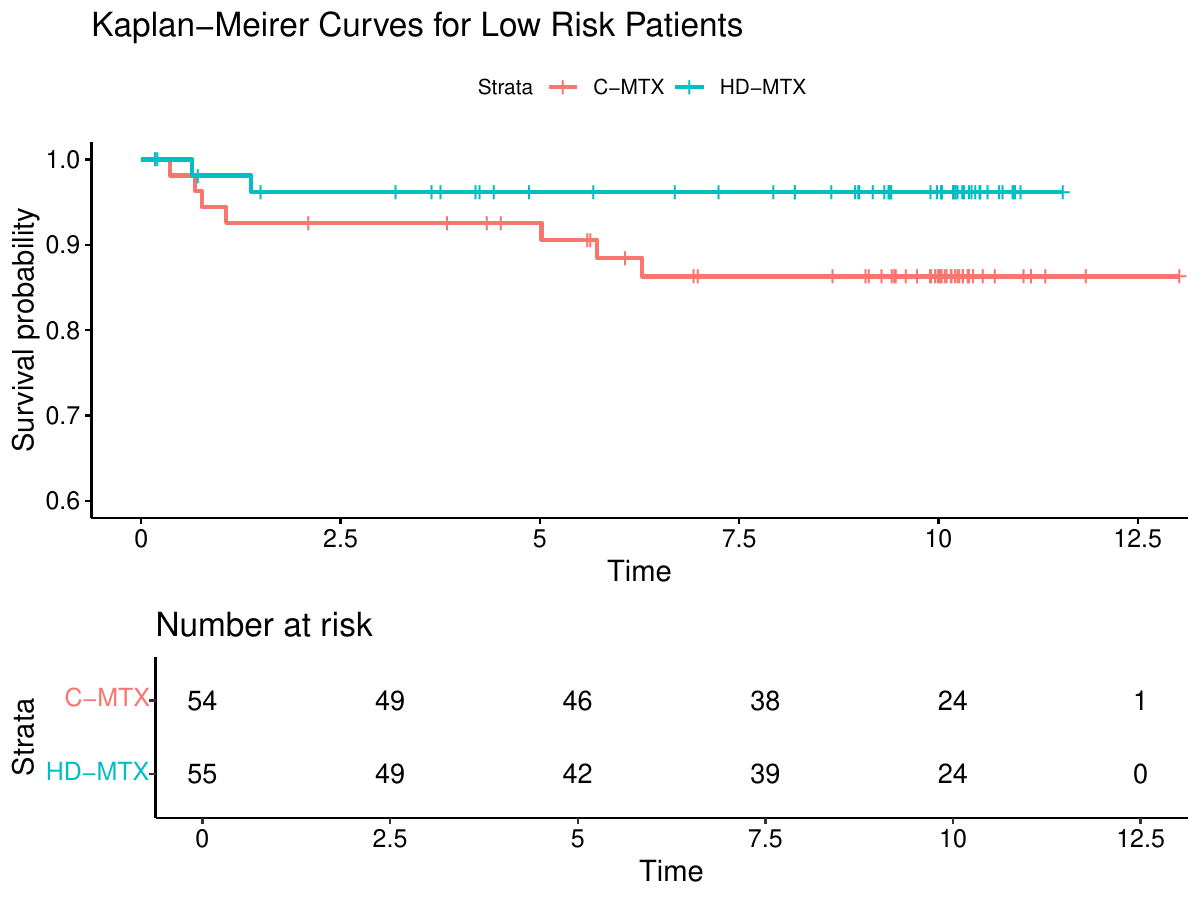}
  \end{subfigure}
  \hfill
  \begin{subfigure}[b]{0.45\textwidth}
    \includegraphics[width=\textwidth]{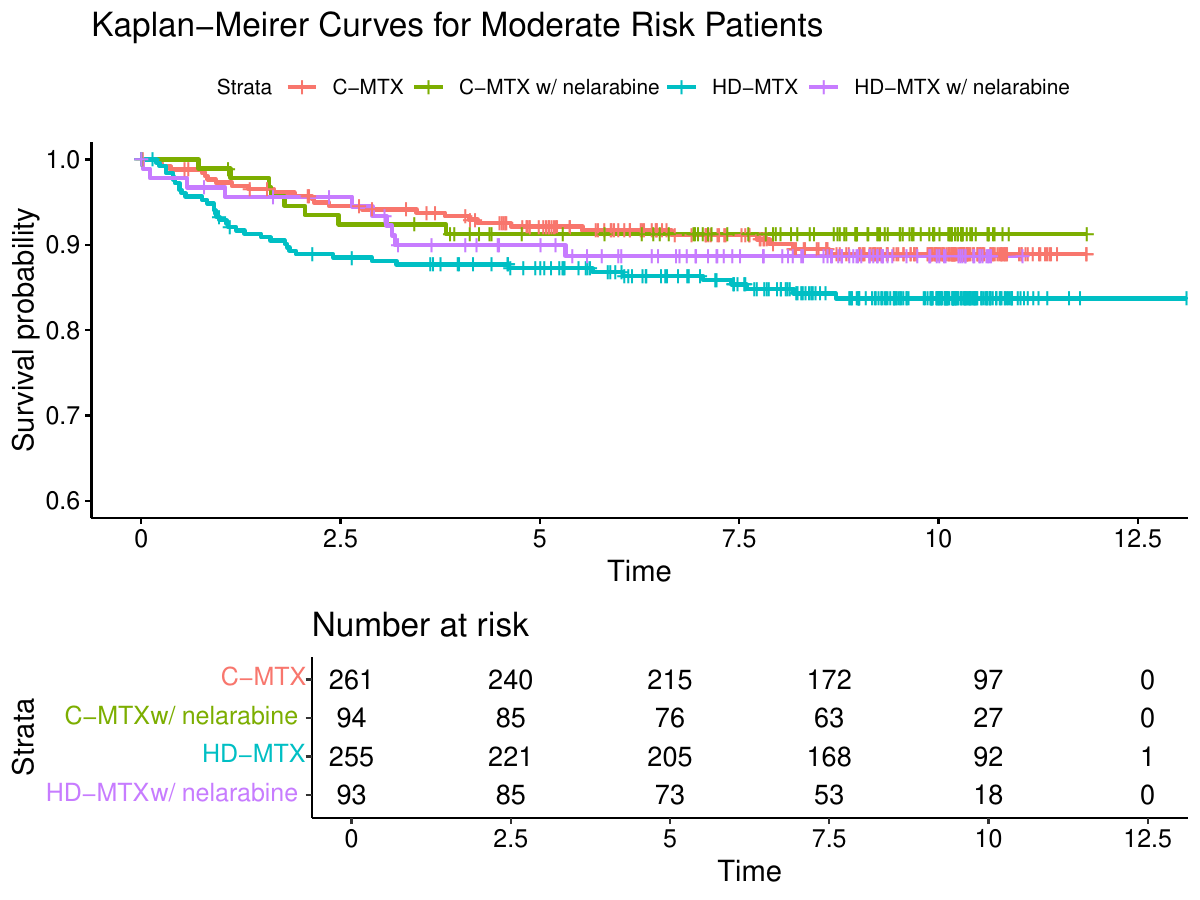}
  \end{subfigure}

  \vskip\baselineskip

  \begin{subfigure}[b]{0.45\textwidth}
    \centering
    \includegraphics[width=\textwidth]{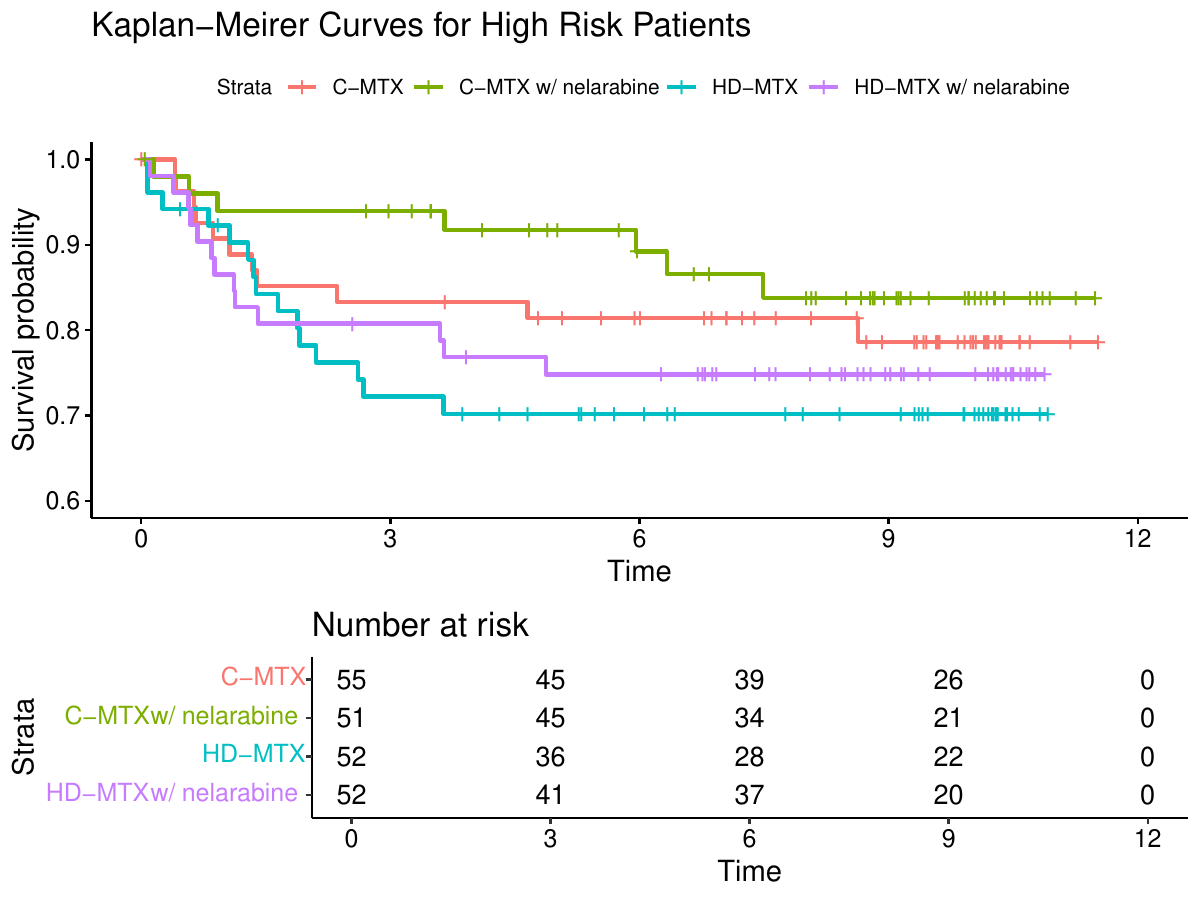}
  \end{subfigure}

  \caption{Kaplan--Meier estimates stratified by risk status.}
  \label{fig:km_risk}
\end{figure}

\newpage
\section{RMST posterior convergence diagnostics}

Gelman-Rubin $\hat R$ and effective sample size (ESS) for the RMST posterior are computed via the \texttt{coda} package using a split-chain approach. We checked across both the linear and nonlinear fits, both cure-rate levels under $\lambda_{g,p} \sim \Unif(-2,0)$ (setting 1), for a representative treatment contrast ($\zeta_{21}$) and, for the nonlinear scenario, $K=4$.


\begin{table}[!htpb]
\centering
\small
\caption{Summary of RMST convergence diagnostics by scenario and cure rate (setting 1).}
\begin{tabular}{llrrr}
\hline
\textbf{Scenario} & \textbf{Cure Rate} & $\hat R$ \textbf{(min/median/max)} & \multicolumn{2}{c}{\textbf{ESS (min/median/max)}} \\
\hline
Linear & High & 1.001 / 1.007 / 1.015 & \multicolumn{2}{c}{214.1 / 448.6 / 896.4} \\
Linear & Low & 1.007 / 1.012 / 1.028 & \multicolumn{2}{c}{167.4 / 188.2 / 560.5} \\
Nonlinear & High & 1.000 / 1.000 / 1.017 & \multicolumn{2}{c}{225.9 / 796.5 / 890.7} \\
Nonlinear & Low & 1.000 / 1.007 / 1.021 & \multicolumn{2}{c}{386.6 / 831.2 / 873.2} \\
\hline
\end{tabular}
\end{table}

Across the combinations checked, split-chain $\hat R$ remained below 1.03
(minimum 0.9996; maximum 1.0283), and ESS exceeded 165 among 1,000
retained draws (median 538.1).

Model fitting was performed on the University of Florida HiPerGator3 high-performance computing cluster, using nodes equipped with AMD EPYC ``Milan'' processors (64 cores, 512 GB RAM per node) connected via InfiniBand. Each of the five candidate values of $K$ was fit as an independent job allocated 4 CPU cores and 50 GB of memory, with 1{,}000 posterior draws retained per fit; peak memory usage across all five fits ranged from approximately 10.2-10.4 GB.
 
Wall-clock runtimes for the nonlinear mixture cure model fit were 2h 20m for $K=6$, 2h 36m for $K=8$, 2h 26m for $K=10$, 2h 58m for $K=15$, and 2h 55m for $K=20$. Since the five fits were run concurrently as an array job, total wall-clock time to complete model fitting across all candidate values of $K$ was approximately 3 hours.

\subsection{AALL0434 data}

For the four AALL0434 treatment-specific RMST posteriors, the diagnostic based on 1,000 retained draws gives $\widehat R$ values from 1.0002 to 1.0062 and full-chain ESS values range from 195.0 to 638.0; the final posterior summaries use all 1,000 retained draws. The trace plots show no evident drift or prolonged sticking. 

\begin{figure}
    \centering
    \includegraphics[width=0.4\linewidth]{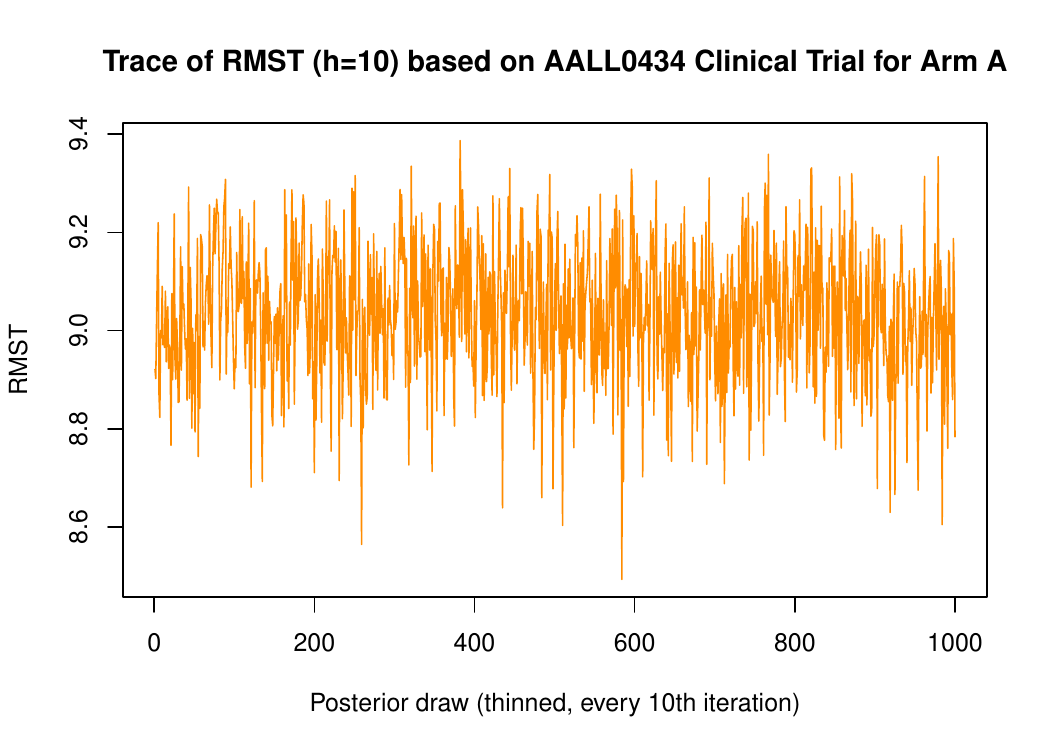}
    \includegraphics[width=0.4\linewidth]{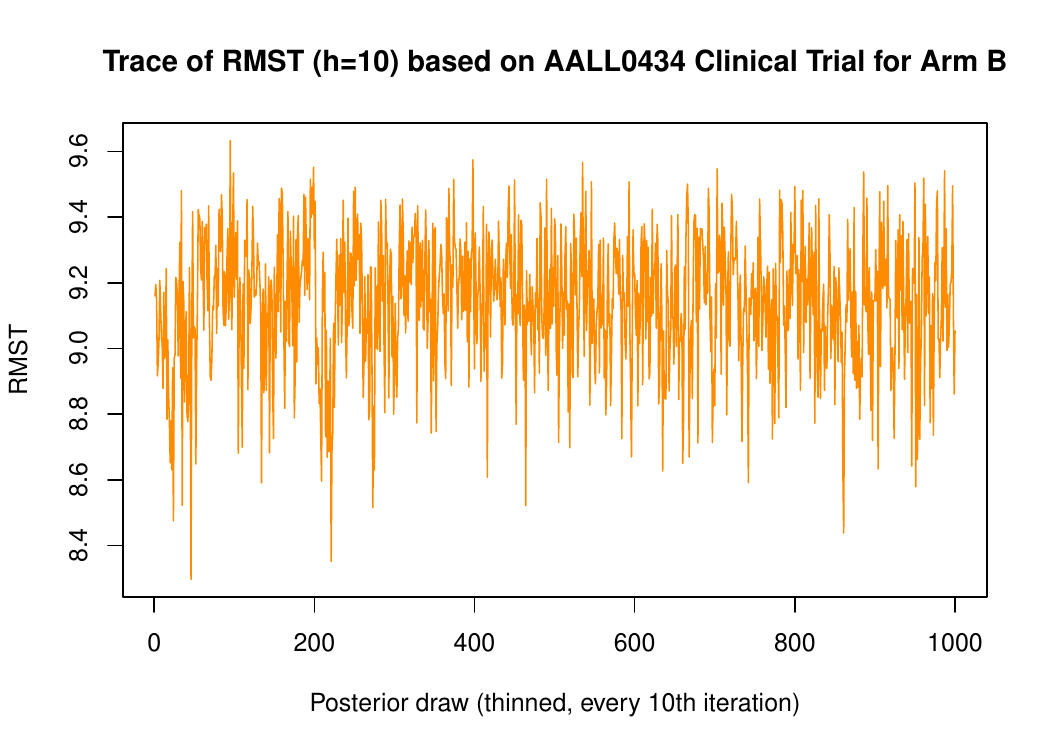}
    \includegraphics[width=0.4\linewidth]{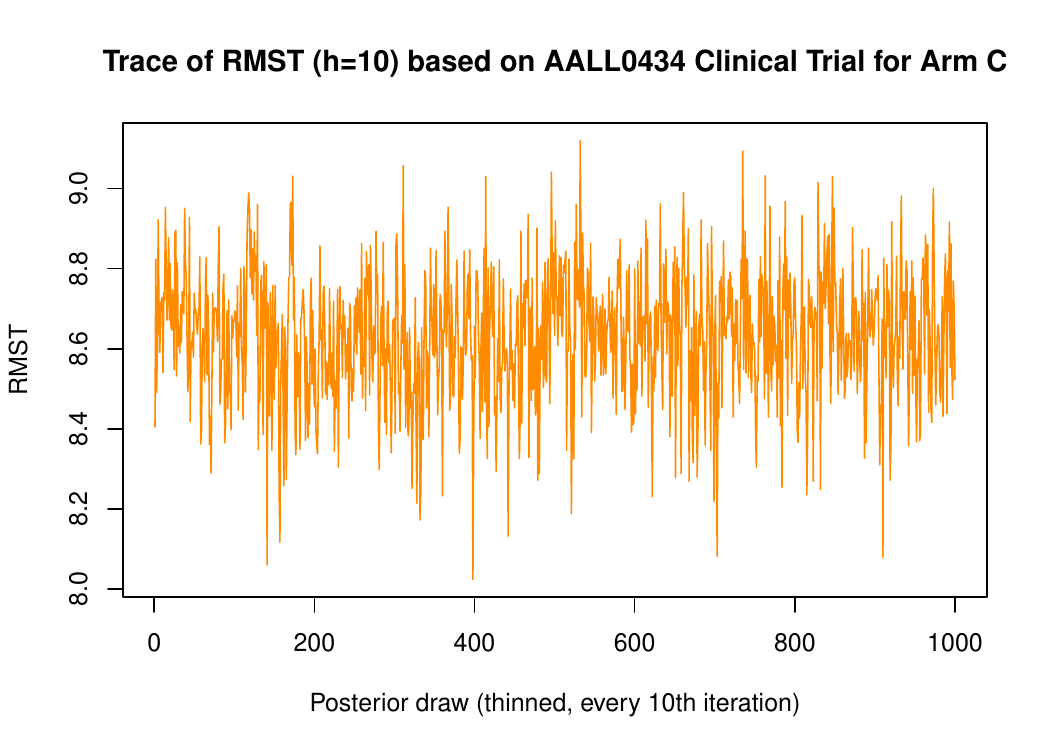}
    \includegraphics[width=0.4\linewidth]{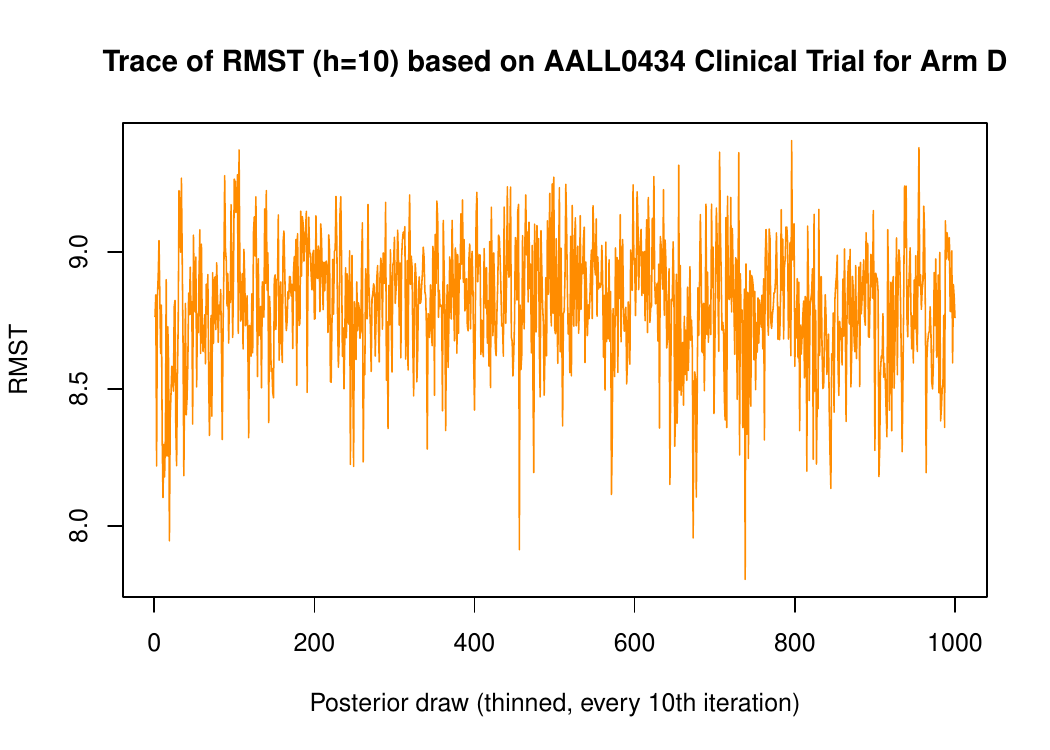}
    \caption{Traceplots of the RMSTs for the four treatment arms in the AALL0434 data, where Arm A = {C-MTX} Arm B = {C-MTX/nelarabine} Arm C = {HDMTX} Arm D = {HDMTX/nelarabine}}
    \label{fig:placeholder}
\end{figure}


\clearpage
\section{Pairwise GRF fits in a multi-treatment setting}

We conduct two diagnostic checks of pairwise \texttt{grf} fits in a
three-treatment setting. Part~A examines whether treatment contrasts
obtained from separate pairwise models are mutually coherent. Part~B
describes how \texttt{grf}'s internal outcome-regression estimate changes
when the same treatment arm is analyzed with different comparator arms.
The two checks address related but distinct questions.

\subsection{Part A: Coherence of pairwise treatment contrasts}

Let $\widehat{\mu}_g(\mathbf{x})$ denote the estimated RMST under treatment
arm $g$, and define
\[
\widehat{\tau}_{ab}(\mathbf{x})
=
\widehat{\mu}_a(\mathbf{x})-\widehat{\mu}_b(\mathbf{x}).
\]
Any three contrasts derived from a common set of treatment-specific RMST
estimates must satisfy
\[
\widehat{\tau}_{32}(\mathbf{x})
=
\widehat{\tau}_{31}(\mathbf{x})
-
\widehat{\tau}_{21}(\mathbf{x}).
\]

Separate pairwise \texttt{grf} fits do not enforce this identity. We
therefore evaluate all three pairwise models on the same held-out test
sample and compare the directly estimated contrast
$\widehat{\tau}_{32}$ with the implied contrast
$\widehat{\tau}_{31}-\widehat{\tau}_{21}$. For each patient, we define
the coherence gap as
\[
\mathrm{gap}
=
(\widehat{\tau}_{31}
-
\widehat{\tau}_{21})
-
\widehat{\tau}_{32},
\]
and the relative coherence gap as
\[
\mathrm{relative\ gap}
=
\frac{
(\widehat{\tau}_{31}
-
\widehat{\tau}_{21})
-
\widehat{\tau}_{32}
}{
\widehat{\tau}_{32}
}.
\]
The absolute gap measures disagreement on the RMST scale. The relative gap
expresses disagreement relative to the directly estimated
$\widehat{\tau}_{32}$.

\begin{table}[!htpb]
\centering
\small
\caption{Difference between the implied contrast
$\widehat{\tau}_{31}-\widehat{\tau}_{21}$ and the directly estimated
contrast $\widehat{\tau}_{32}$ in one representative replication.}
\begin{tabular}{lccccc}
\hline
\textbf{Scenario} &
\textbf{Median $|$gap$|$} &
\textbf{Mean $|\widehat{\tau}_{32}|$} &
\textbf{Median $|$rel.\ gap$|$} \\
\hline
Well-specified    & 0.4151 & 3.9209 & 0.1059 \\
Misspecified  & 0.0722 & 0.6565  & 0.1027 \\
\hline
\end{tabular}
\end{table}

The median absolute relative gap is 10.6\% in the linear scenario and
10.3\% in the nonlinear scenario. In this representative replication, the
direct and implied contrasts therefore differ materially; the absolute-gap
results avoid overinterpreting ratios for subjects whose direct contrast is
near zero.

\subsection{Part B: Sensitivity of internal outcome estimates to the
pairwise fitting context}

Each treatment arm appears in two pairwise \texttt{grf} fits. For
patients in a given arm, we retain \texttt{Y.hat} from both fits. In
\texttt{grf}, \texttt{Y.hat} is the cross-fitted, out-of-bag estimate of
the conditional outcome mean used as a nuisance function during model
fitting. It is not a treatment-specific potential-outcome or survival-
function estimate.

For each patient, we calculate the absolute difference
\[
\mathrm{diff}
=
\left|
\widehat{Y}^{(1)}
-
\widehat{Y}^{(2)}
\right|,
\]
where the superscripts identify the two pairwise training samples that
contain the patient's treatment arm. The patient, covariates, and
observed outcome are unchanged; only the comparator arm, and therefore
the pairwise training sample, differs. We also calculate the signed
relative difference
\[
\mathrm{relative\ diff}
=
\frac{
\widehat{Y}^{(1)}
-
\widehat{Y}^{(2)}
}{
\widehat{Y}^{(1)}
},
\]
using the first context as an arbitrary reference. Unlike Part~A, which
uses predictions on a held-out test sample, this check uses out-of-bag
estimates for observations included in the corresponding training
samples.

\begin{table}[!htpb]
\centering
\scriptsize
\setlength{\tabcolsep}{4pt}
\caption{Variation in \texttt{grf}'s out-of-bag outcome-regression
estimates across the two pairwise training contexts containing each
treatment arm. Results are from one representative replication.}
\begin{tabular}{llrrrrrrr}
\hline
\textbf{Scenario} &
\textbf{Arm} &
\textbf{N} &
\textbf{Mean $|$diff$|$} &
\textbf{Median $|$diff$|$} &
\textbf{Max $|$diff$|$} &
\textbf{Mean rel.\ diff} &
\textbf{Median $|$rel.\ diff$|$} \\
\hline
Well-specified    & 1 & 45 & 0.0872 & 0.0745 & 0.1537 & 0.0992    & 0.0843  \\
Well-specified    & 2 & 34 & 0.0176 & 0.0191 & 0.0311 & $-0.0199$ & 0.0220  \\
Well-specified    & 3 & 41 & 0.1100 & 0.1031 & 0.1504 & $-0.1427$ & 0.1311  \\
Misspecified  & 1 & 36 & 0.0755 & 0.0669 & 0.1770 & 0.0918    & 0.0818  \\
Misspecified & 2 & 43 & 0.0900 & 0.0888 & 0.1377 & 0.1089    & 0.1070  \\
Misspecified & 3 & 41 & 0.0217 & 0.0215 & 0.0605 & 0.0237    & 0.0276 \\
\hline
\end{tabular}
\end{table}

The amount of sensitivity differs across arms and scenarios. These results indicate that the internal nuisance estimate can
change appreciably when the comparator arm and training sample change.
Since the estimated survival function is a nuisance estimate whose target can depend on
the pairwise sample, a difference between the two values does not by
itself indicate an invalid fit. The direct evidence about incompatibility
among the estimated treatment contrasts is provided by Part~A. Together,
the checks show that separate pairwise fits may depend on the chosen
comparison and, unlike a joint multi-treatment model, do not enforce a
common set of mutually coherent treatment contrasts.

\clearpage

\section{The advantage of shared learning over T-learning}

In this 150-replication simulation, we examine whether updating $\bGamma$
improves accuracy and reduces posterior uncertainty. Every model starts
with a block-diagonal $\bGamma$, so each treatment arm initially has its own
mixture components. In the sharing model, $\bGamma$ is updated during
MCMC, allowing treatments to share components. In the no-sharing model,
$\bGamma$ remains fixed, producing T-learning-type estimates based on
separate component sets. We generate data from the well-specified model and
compare linear-link models under both cure-rate settings. A comparison based on the misspecified model was also considered, but it was not pursued due to the difficulty in isolating the effects of component sharing and model approximation. MSE is the error
in the estimated RMST treatment contrast. Posterior uncertainty is
summarized by the arm-specific RMST posterior variance, averaged across
replications.

\begin{table}[!htpb]
\centering
\small
\caption{MSE (mean squared error vs.\ truth) for RMST contrasts, $\Gamma$-updated vs.\ $\Gamma$-fixed.}
\begin{tabular}{llrr}
\hline
\textbf{Scenario} & \textbf{Contrast} & \textbf{$\Gamma$-updated} & \textbf{$\Gamma$-fixed} \\
\hline
High cure rate   & $\zeta_{21}$ & 3.10 & 3.76 \\
                         & $\zeta_{31}$ & 1.02 & 1.58 \\
                         & $\zeta_{32}$ & 2.77 & 5.67 \\
\hline
Low cure rate    & $\zeta_{21}$ & 14.33 & 18.21 \\
                         & $\zeta_{31}$ & 1.49 & 2.01 \\
                         & $\zeta_{32}$ & 14.23 & 19.57 \\
\hline
\end{tabular}
\end{table}

\begin{table}[!htpb]
\centering
\small
\caption{Posterior variance of RMST by arm, mean across replications, $\Gamma$-updated vs.\ $\Gamma$-fixed.}
\begin{tabular}{llrr}
\hline
\textbf{Scenario} & \textbf{Arm} & \textbf{$\Gamma$-updated} & \textbf{$\Gamma$-fixed} \\
\hline
High cure rate   & Arm 1 & 0.2417 & 0.9676 \\
                         & Arm 2 & 0.1859 & 0.2444 \\
                         & Arm 3 & 0.2446 & 1.1862 \\
\hline
Low cure rate    & Arm 1 & 0.7775 & 2.0380 \\
                         & Arm 2 & 0.4404 & 0.3569 \\
                         & Arm 3 & 0.6492 & 2.0088 \\
\hline
\end{tabular}
\end{table}

Updating $\bGamma$ reduces the MSE in all six scenario--contrast
combinations.
Updating $\bGamma$ also reduces the posterior variance in five of the
six arm--scenario combinations, often substantially. The exception is
Arm~2 under the low cure rate, for which the posterior variance is
slightly larger under sharing.

In these simulations, all treatment arms share at least one active
component, and some arms share additional components. The updated
$\bGamma$ can use this common structure while retaining components that
are specific to individual arms. In contrast, fixing $\bGamma$ at a
block-diagonal specification prevents any borrowing across treatments.
Overall, the results show that adaptive sharing can improve accuracy and
generally reduce posterior uncertainty when the treatment-specific
survival distributions have common features.

\clearpage
\baselineskip=12pt
\bibliographystyle{bibstyle}
\bibliography{mixture_rev}
\end{document}